%% file: master.tex
\documentclass{article}

\usepackage{arxiv}

\usepackage{amsmath,amsthm,amssymb,commath,bm,mathrsfs}
\usepackage{physics}
\usepackage[normalem]{ulem}
\usepackage{listings,color,graphicx}
\usepackage{tabularx}
\usepackage{float}
\usepackage{subfloat}
\usepackage{subfig}
\usepackage{xspace}
\usepackage{fixltx2e}
\usepackage{booktabs}
\usepackage{hyperref}
\usepackage{algpseudocode}
\usepackage{algorithm}
\usepackage{placeins}
\usepackage[utf8]{inputenc} 
\usepackage[T1]{fontenc}
\usepackage{makecell}
\usepackage{multirow}
\usepackage{cases}
\usepackage{array}

\newcommand{\AlgAbb}{\texttt{SPASD}}    % Algorithm Abbreviation
\newcommand{\FPROG}{\mathcal{F}}
\newcommand{\CPROG}{\mathcal{G}}

\newcommand{\LDS}{CS}   % low-dimensional solver
\newcommand{\HDS}{FS}
\newcommand\blfootnote[1]{%
  \begingroup
  \renewcommand\thefootnote{}\footnote{#1}%
  \addtocounter{footnote}{-1}%
  \endgroup
}

\title{Multiscale parareal algorithm for long-time mesoscopic simulations of microvascular blood flow in zebrafish}

\author{
    Ansel L. Blumers$^*$\\
    Department of Physics\\ 
    Brown University\\ 
    Providence, RI 02912, USA\\
    \And
    Minglang Yin$^{*}$\\
    Center for Biomedical Engineering\\
    School of Engineering\\
    Brown University\\
    Providence, RI 02912, USA\\
    \And
    Hiroyuki Nakajima\\
    Department of Cell Biology\\
    National Cerebral and Cardiovascular Center Research Institute\\
    Osaka, Japan\\
    \And
    Yosuke Hasegawa\\
    Institute of Industrial Science\\
    The University of Tokyo\\
    Tokyo, Japan\\
    \And
    Zhen Li\\
    Department of Mechanical Engineering\\
    Clemson University\\
    Clemson, SC 29634, USA\\
    \And
    George Em Karniadakis\\
    Division of Applied Mathematics\\
    Brown University\\
    Providence, RI 02912, USA\\
}

\begin{document}
\maketitle

\blfootnote{{$^*$ These authors contributed equally to this work.}}

\begin{abstract}
Various biological processes such as transport of oxygen and nutrients, thrombus formation, vascular angiogenesis and remodeling are related to cellular/subcellular level biological processes, where mesoscopic simulations resolving detailed cell dynamics provide a key to understanding and identifying the cellular basis of disease. However, the intrinsic stochastic effects can play an important role in mesoscopic processes, while the time step allowed in a mesoscopic simulation is restricted by rapid cellular/subcellular dynamic processes. These challenges significantly limit the timescale that can be reached by mesoscopic simulations even with high-performance computing. To break this bottleneck and achieve a biologically meaningful timescale, we propose a multiscale parareal algorithm in which a continuum-based solver supervises a mesoscopic simulation in the time-domain. Using an iterative prediction-correction strategy, the parallel-in-time mesoscopic simulation supervised by its continuum-based counterpart can  converge fast. The effectiveness of the proposed method is first verified in a time-dependent flow with a sinusoidal flowrate through a Y-shaped bifurcation channel. The results show that the supervised mesoscopic simulations of both Newtonian fluids and non-Newtonian bloods converge to reference solutions after a few iterations. Physical quantities of interest including velocity, wall shear stress and flowrate are computed to compare against those of reference solutions, showing a less than 1\% relative error on flowrate in the Newtonian flow and a less than 3\% relative error in the non-Newtonian blood flow. The proposed method is then applied to a large-scale mesoscopic simulation of microvessel blood flow in a zebrafish hindbrain for temporal acceleration. The three-dimensional geometry of the vasculature is constructed directly from the images of live zebrafish under a confocal microscope, resulting in a complex vascular network with 95 branches and 57 bifurcations. The time-dependent blood flow from heartbeats in this realistic vascular network of zebrafish hindbrain is simulated using dissipative particle dynamics as the mesoscopic model, which is supervised by a one-dimensional blood flow model (continuum-based model) in multiple temporal sub-domains. The computational analysis shows that the resulting microvessel blood flow converges to the reference solution after only two iterations. The proposed method is suitable for long-time mesoscopic simulations with complex fluids and geometries. It can be readily combined with classical spatial decomposition for further acceleration.

\end{abstract}

\keywords{multiscale modeling, zebrafish, vascular network, parallel-in-time, dissipative particle dynamics, 1D blood flow modeling}

%%%%%%%%%%%%%%%%%%%%%%%%%%%%%%%%%%%%%%%%%%%%%%%%%%%%%%%%%%%%%%%%%%%%%%%%%%%%%%%%%%%%%%%%%%%%%%%%%%%%
\section{Introduction} \label{sec:intro}
Mesoscopic simulations are becoming increasingly important on studying material properties and interesting phenomena at the mesoscale, i.e., in biological and physical systems, thanks to its ability to reveal detailed dynamics with flexible timescales. As the spatiotemporal scales become small, the intrinsic stochastic effects can play important role in mesoscopic processes, which makes the continuum hypothesis become questionable. Instead, the mesoscopic system dynamics is suitably described by Lagrangian particle methods based on atomistic and molecular models~\cite{monaghan2012smoothed, groot1997dissipative, espanol1995statistical}. Dissipative particle dynamics (DPD), as one of the most promising models in mesoscopic simulations, is a bottom-up coarse-grained particle method, derived directly from the molecular dynamic formulation, and governed by the Newton's law with calibrated coarse-grained force field. DPD and its variants have been widely applied to polymer science~\cite{ symeonidis2005dissipative, 2019Wang_Implicit}, biofluids~\cite{blumers2017gpu, Pivkin2008}, and hydrodynamics~\cite{2018Bian_ANote, 2020Xia_AGPU}, to name but a few. Thermal fluctuations and their correlations at mesoscale are important and appear as  a stochastic term in the governing equations. To fully resolve molecular details and capture the underlying dynamics at the mesoscopic level, a time step at the order of a picosecond is needed~\cite{2019Wang_Concurrent}, which greatly limits the potential of mesoscopic models to simulate {\it in vivo} biological or physical processes with a larger characteristic time scale.

One promising strategy to overcome such bottleneck is by leveraging parallel computing. Following the philosophy of ``divide and conquer", parallel computing divides the whole spatial domain of the system into multiple subsystems, where each subsystem will be simulated simultaneously on distributed computing nodes. However, the computing efficiency may be limited when the number of distributing cores are too many to efficiently communicate in a message passing scheme~\cite{blumers2017gpu}. As an alternative, a parallel-in-time strategy known as ``parareal'' originally proposed by Lions, Maday and Turinici~\cite{Lions2001} has been widely accepted as a parallel-in-time integration method. In the parareal method, a single long-time simulation is divided into multiple consecutive time-domains and distributed onto $n$ parallel simulations for representing the system dynamics at time $t_{1}, t_{2}, ..., t_{n}$. Essentially, parareal adopts an iterative prediction-correction strategy, where each subsystem at different time will communicate between its neighbors in time and make prediction with correction until convergence~\cite{gander2007analysis}. Following the idea of the parareal, Blumers et al.~\cite{Blumers2019} proposed a supervised parallel-in-time algorithm for stochastic dynamics (~\AlgAbb{}), which couples the finite element (FEM) solver as a supervisor with a DPD solver. Following the guidance of the FEM solver, the~\AlgAbb{} method converges (exponentially) fast after a number of iterations.
%and demonstrates its superiority when the simulation requires a large number of cores.

To illustrate how the multiscale parareal method can facilitate realistic simulations and reveal detailed dynamics, we consider an interesting biology model, the zebrafish, as our experimental model and simulate its hemodynamics in the hindbrain. Zebrafish has been receiving an increasing amount of attention in fields such as cancer biology, vascular angiogenesis and remodeling. Specifically, vascular remodeling is a biological process, where some unnecessary vessels in pri-mature and unstructured vascular networks formed from angiogenesis~\cite{Risau1997} are subsequently pruned to improve the transport properties of the entire vasculature. As a result, an efficient matured network characterized by a hierarchical structure is achieved~\cite{Mirzapour-shafiyi2020}. Although angiogenesis is considered to be mainly driven by chemical signals secreted from hypoxic tissues, 
recent evidence has shown that hemodynamics plays also an important role in vessel pruning~\cite{Bernabeu2014,Huang2003}. Shear stress on vessel wall stimulates vessel regression through chemical signaling~\cite{Franke1984,Resnick1995}, which is followed by retraction, apoptosis, and reintegration of endothelial cells. Due to the difficulty of estimating the force acting on the vessel wall, however, the relationship between the hemodynamic parameters and the vessel pruning has not been fully clarified. Zebrafish possesses a unique feature that its embryo is transparent, so that \textit{in-vivo} imaging of vessel structures and blood cell dynamics is possible. In the present study, we propose to apply the multiscale parareal algorithm on simulating the blood flow in zebrafish hindbrain, where such simulation is restricted by the expensive computational cost and efficiency. To the best of the authors' knowledge, this is the first attempt to clarify the hemodynamic parameters through detailed three-dimensional (3D) simulation of the zebrafish hindbrain. 

Using the continuum formulation, the computational cost for solving 3D Navier-Stokes(NS) equations could be extremely high since a fine mesh is needed to discretize such a complex 3D spatial domain. Instead, a one-dimensional (1D) blood flow model is adopted to supervise the expensive mesoscopic simulator as it predicts vessel blood flow at satisfactory accuracy with a low computational cost~\cite{sherwin2003one, formaggia2003one}. The complex vascular network in the zebrafish hindbrain is represented as a 1D pipe-line network where vessel branches merge and diverge at bifurcation points. Along a single branch, the velocity and area at grid points are computed to characterize the local hemodynamics. The 1D model can be derived from the full 3D Navier-Stokes equations under the assumptions of incompressibility, axial dominance in velocity, and the absence of turbulence~\cite{sherwin2003one}. Many successful biological and biomedical applications have proven the versatility and low computational cost of the 1D blood flow model in predicting blood flow and pressure in coronary arteries\cite{mynard20081d, yin2019one}, circle of Willis~\cite{grinberg2011modeling}, or abdomial arteries~\cite{sherwin2003computational, raghu2011comparative}. Therefore, we adopt the 1D blood flow model in our continuum-based solver in this study.

We apply the multiscale parareal algorithm for accelerating massive Lagrangian simulations of blood flow at mesoscale to zebrafish hindbrain, see Fig.~\ref{fig:hf_lf}. We will also demonstrate that further computational speedup can be achieved by combining the present strategy with a classical spatial decomposition method. The reminder of this paper is organized as follows: we first introduce~\AlgAbb{} for the application of blood flow in Section~\ref{sec:solvers} and~\ref{sec:SPASD}. We then explain in detail the coupling scheme for on-the-fly communications between our coarse and fine solvers (CS/FS) in Section~\ref{sec:multiframework}, containing the algorithmic architecture that enables real-time data transfer in massive parallel simulations. In Section~\ref{sec:numericalresults}, we present quantitative results of mesoscopic simulations accelerated by our~\AlgAbb{} framework. Finally, Section~\ref{sec:discussion} contains a brief summary and discussion. 

%%%%%%%%%%%%%%%%%%%%%%%%%%%%%%%%%%%%%%%%%%%%%%%%%%%%%%%%%%%%%%%%%%%%%%%%%%%%%%%%%%%%%%%%%%%%%%%%%%%%
\section{Algorithm} 

%%%%%%%%%%%%%%%%%%%%%%%%%%%%%%%%%%%%%%%%%%%%%%%%
\subsection{Methods} \label{sec:solvers}

\begin{figure}
    \centering
	\includegraphics[width=0.8\textwidth]{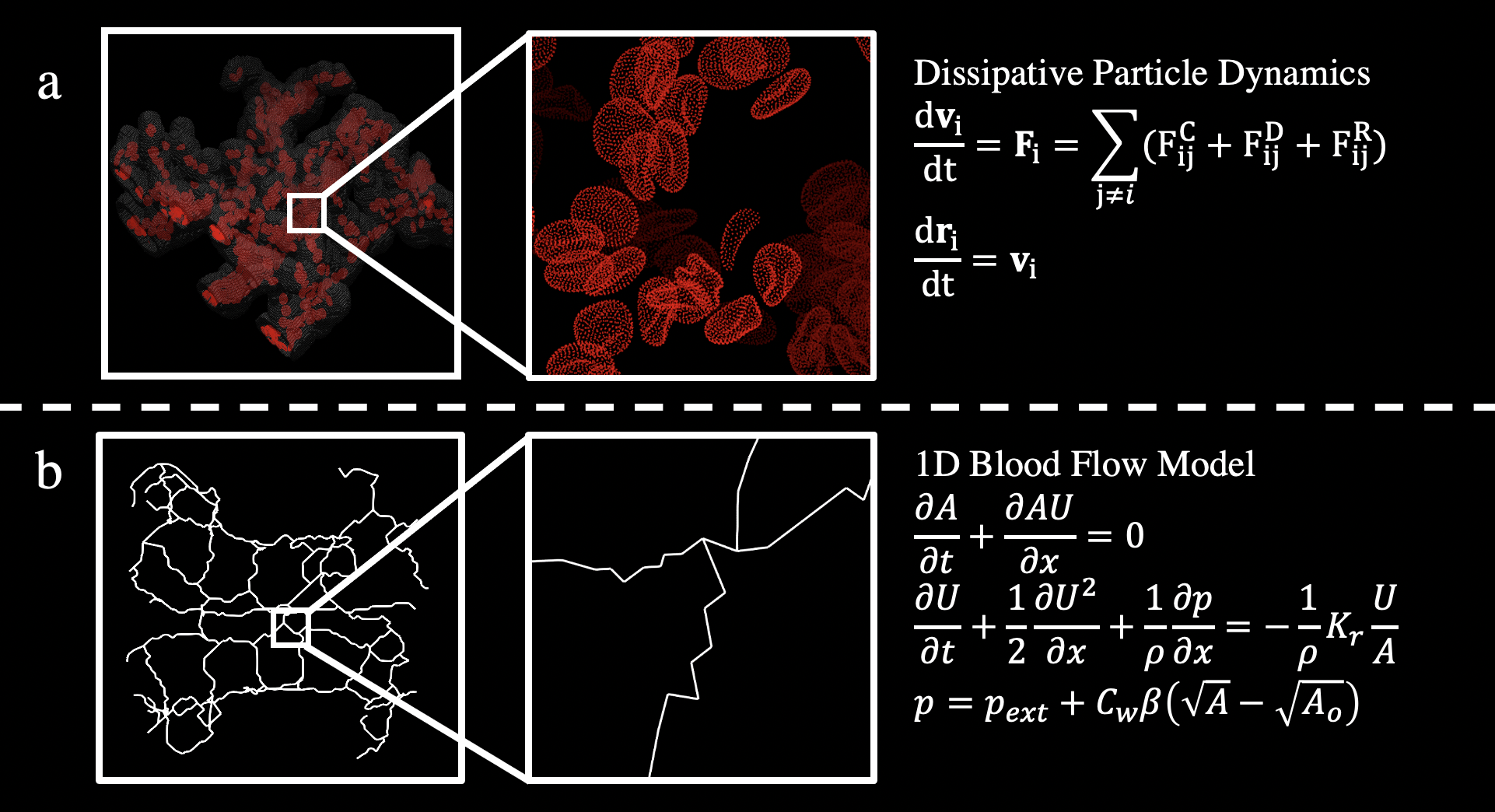} 
    \caption{\textbf{Overview of the (a) DPD model (FS) and the (b) 1D model (CS).} (a) The DPD model describes the detailed blood flow using discrete particles and RBCs made up of DPD particles. (b) The 1D model represents a macroscopic view of the system, where the 3D vasculature can be simplified as a pipeline system composing of elastic pipes. The governing equations for both models are presented on the right.}
    \label{fig:hf_lf}
\end{figure}

\begin{table}[]
\centering
\begin{tabular}{|l|l|l|}
\hline
                        & \textbf{Coarse Solver (CS)} & \textbf{Fine Solver (FS)}      \\ \hline
%\textbf{Name}           & NEKTAR-1D              & \USERMESO{} 2.0                     \\ \hline
\textbf{Type}           & Continuum model       & Particle model                     \\ \hline
\textbf{Method}         & 1D Blood Flow Model & DPD \\ \hline
\textbf{Dimensionality} & One-dimension          & Three-dimension                     \\ \hline
\textbf{Code}           & Serial                 & Parallel                            \\ \hline
\textbf{Hardware support} & CPU                  & GPU                                 \\ \hline
\end{tabular}
\caption{ Features of the coarse solver and fine solver in \AlgAbb.}
\label{tab:coarse_fine}
\end{table}

The high-fidelity Lagrangian simulator of \AlgAbb{} employs DPD, which is referred as the "fine solver" (FS, as shown in Fig. \ref{fig:hf_lf}). DPD is a widely used particle model for mesoscopic simulation of complex fluids, where each DPD particle is represented explicitly by its position and velocity. This approach offers a bottom-up approach that resolves motion with Newton's equation and provides a flexible representation of various types of fluid. In particular, Pivkin et al.~\cite{Pivkin2008} developed a mixture of DPD particles and coarse-grained mesh-based cells to simulate the motion of red blood cells with ambient flows in vessels. In the multiscale parareal algorithm, FS is complemented by a low-fidelity 1D model for simulating blood flow in arteries (referred as the "coarse solver", CS). The 1D model provides a macroscopic yet less detailed description of blood flow by assuming the dominance of blood flow in the axial direction and no perfusion. The 1D solver for blood flows can be executed on a single CPU with a low cost, and these results will then be transferred to DPD models as a guidance in each temporal sub-domain. Table~\ref{tab:coarse_fine} shows a comparison between the two models. More details about the DPD and 1D models are provided in the Appendix.

%%%%%%%%%%%%%%%%%%%%%%%%%%%%%%%%%%%%%%%%%%%%%%%%
\subsection{Supervised Parallel-in-time Algorithm for Stochastic Dynamics (\AlgAbb{})} \label{sec:SPASD}

In this section, we introduce the details of $\AlgAbb$. Notice that it is flexible to choose any combination of a stochastic particle model and its continuum counterpart other than the DPD/1D models~\cite{Blumers2019} for \AlgAbb (Algorithm \ref{alg:zebra_MultiParareal}). Such flexibility is warranted by the deliberate choice of the flow variable $Q_{k}^{n}$, which servers as the medium between meso- and macro-scales. An appropriate flow variable must be easy-to-extract and accurately describe the state of the fluid system. In this work, we take the flowrate along the local axial direction as the flow variable $Q_{k}^{n}$ to couple the \LDS{} and \HDS{} models for the reason that its evolution describes the state of blood flow over time; $k$ denotes the number of iteration and $n$ is the index of temporal sub-domain. For the particle model, the axial flowrate can be estimated from the product of averaged particles velocity on a cross-sectional plane and its cross-sectional area.

%It can be easily tuned by scaling up or down the component in the axial direction while affixing the other components. In general, another candidate for state variable is the axial velocity to communicate between \LDS{} and \HDS{}. However, initializing the DPD system from the 1D axial velocity would not preserve the flow profile. }

\begin{table}[]
\centering
\begin{tabular}{ p{1cm}  p{6cm} }
\hline \hline
$\mathcal{R}$   &       Mapping operator: from CS to FS \\
$\mathcal{P}$   &       Projection operator: from FS to CS\\
$\mathcal{D}$   &       Filtering operator: noise filtering \\
$\CPROG$        &       Coarse propagator: advancing in CS \\
$\FPROG$        &       Fine propagator: advancing in FS \\
$K$             &       Number of iterations\\
$N$             &       Number of time subdomains \\
\hline \hline
\end{tabular}
\caption{Notation for selected variables.}
\label{tab:notation}
\end{table}

\begin{algorithm}
\caption{\AlgAbb{}}
\begin{algorithmic}[1]
\State Initialization: Advance in CS: $\CPROG^{n+1}_{0} = \CPROG(Q_{0}^{n})$ for $0\leq n \leq N-1$.

\State Main loop

\textbf{for} $k = 1,...,K$ \textbf{do}

\hspace{\algorithmicindent} \textbf{DO\_PARALLEL} for $k-1 \le n \le N-1$.

\hspace{\algorithmicindent}\hspace{\algorithmicindent} \textbf{Correction}:

\hspace{\algorithmicindent}\hspace{\algorithmicindent} Filter solution to remove noise: $\mathcal{D}\{Q^n_k\}$ 

\hspace{\algorithmicindent}\hspace{\algorithmicindent} Advance with \LDS{}: $\CPROG^{n+1}_{k} = \CPROG(\mathcal{D}\{Q^n_k\})$ 

\hspace{\algorithmicindent}\hspace{\algorithmicindent} Map the macroscopic state to microscopic state:  $\mathcal{R}\{Q^n_k\}$. 

\hspace{\algorithmicindent}\hspace{\algorithmicindent}\hspace{\algorithmicindent} Skip mapping if $Q^n_k -Q^n_{k-1} < \Delta_{\text{th}}$ where $\Delta_{\text{th}}$ a user-defined threshold.

\hspace{\algorithmicindent}\hspace{\algorithmicindent} Advance with \HDS{}:  $\FPROG(\mathcal{R}\{Q^n_k\})$.

\hspace{\algorithmicindent}\hspace{\algorithmicindent} Project the microscopic state to macroscopic state:  $\mathcal{P}^{n+1}_{k} = \mathcal{P}\{\FPROG(\mathcal{R}\{Q^n_k\})\}$.

\hspace{\algorithmicindent}\hspace{\algorithmicindent} Compute the correction: $\delta^{n+1}_k \equiv \mathcal{P}^{n+1}_{k} - \CPROG^{n+1}_{k}$.

\hspace{\algorithmicindent} \textbf{END\_PARALLEL}

\hspace{\algorithmicindent} \textbf{for} $n=k-1,...,N-1$ \textbf{do}

\hspace{\algorithmicindent}\hspace{\algorithmicindent} \textbf{Prediction}:

\hspace{\algorithmicindent}\hspace{\algorithmicindent} \textbf{if $n = k-1$ do}

\hspace{\algorithmicindent}\hspace{\algorithmicindent} \hspace{\algorithmicindent} Update: $Q^{n+1}_{k+1} = \mathcal{P}^{n+1}_{k}$

\hspace{\algorithmicindent}\hspace{\algorithmicindent} \textbf{else do}

\hspace{\algorithmicindent}\hspace{\algorithmicindent} \hspace{\algorithmicindent} Advance with \LDS{} in serial:  $\CPROG^{n+1}_{c,k} = \CPROG(\mathcal{D}\{Q^{n}_{k+1}\})$.

\hspace{\algorithmicindent}\hspace{\algorithmicindent} \hspace{\algorithmicindent} \textbf{Refinement}:

\hspace{\algorithmicindent}\hspace{\algorithmicindent} \hspace{\algorithmicindent} Combine the correction and the prediction terms: $Q^{n+1}_{k+1} = \CPROG^{n+1}_{c,k} + \delta^{n+1}_{k}$ .

\hspace{\algorithmicindent}\hspace{\algorithmicindent} \textbf{end if}

\hspace{\algorithmicindent}\hspace{\algorithmicindent} 

\hspace{\algorithmicindent} \textbf{end}

\hspace{\algorithmicindent} Break if the solution converges.

\textbf{end}

\end{algorithmic}
\label{alg:zebra_MultiParareal}
\end{algorithm}

\begin{figure} 
\centering
\includegraphics[width=1\textwidth]{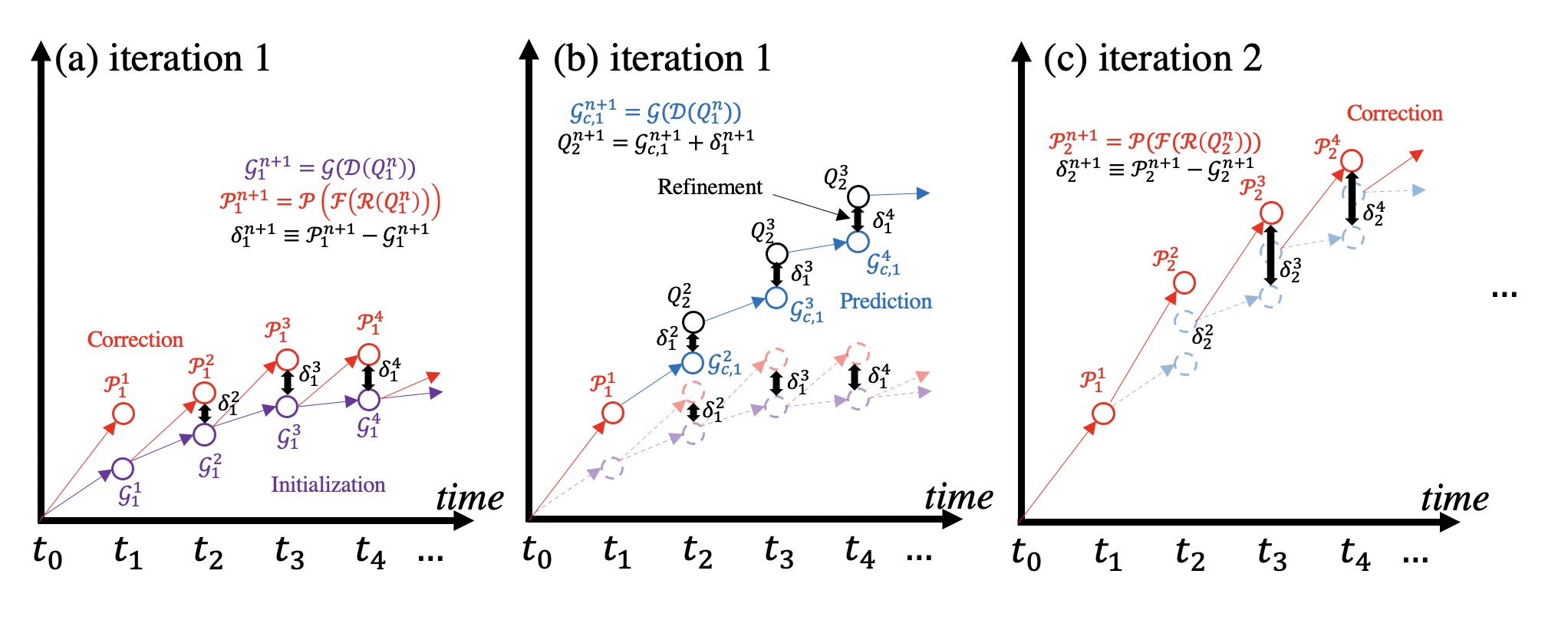}
\caption{\textbf{A graphical representation of $\AlgAbb$.} The flow variable $Q^{n}_{k}$ is computed via correction, prediction, and  refinement in each iteration. After initialization, FS $\mathcal{P}$ will advance the system in parallel within its temporal sub-domain, encompassing other operations, i.e., mapping $\mathcal{R}$, projection $\mathcal{P}$, and $\mathcal{D}$. The correction terms $\delta$ are computed at the end of each temporal sub-domain to correct the prediction in a serial propagation \LDS{} in prediction and refinement. (a) and (b) present the first iteration. (c) shows the second iteration starting from correction.}
\label{fig:graphic}
\end{figure}

\AlgAbb{} is summarized in Algorithm \ref{alg:zebra_MultiParareal} and represented graphically in Fig. \ref{fig:graphic}. \LDS{} advances independently in the initialization. Since the governing equations of the macroscopic model are very different from those of the microscopic model, the form and dimension of the solution representing the macroscopic state substantially varies from those representing the microscopic state. The solution from \LDS{} can be mapped to a microscopic state via a mapping operation (denoted by mapping operator $\mathcal{R}$) while the converse mapping from \HDS{} to \LDS{} can be obtained by the projection operator $\mathcal{P}$. Typical methods such as weighted kernel sampling and interpolation are employed for the mapping and projection operations. In the current application, the flowrate is computed at centerline points, which are shared by both models because the 1D geometry is extracted from the 3D geometry. Therefore, the mapping and projection operations are not required here. Moreover, noise filtering $\mathcal{D}$ is essential to the success of \AlgAbb{} because the noise from the \HDS{} would accumulate with time, which will subsequently lead to a divergence of refined solution from the true solution if left unattended. In this work, noise filtering is embedded in the advancement of \LDS{}. 
%\textcolor{red}{XXX MY: Why? Why first order leads to noise canceling? "The governing equations of \LDS{} exhibit strong noise smoothing characteristics as the spatial derivative on flowrate is only first order. Coupled with fixed boundary values, the system can effectively eliminate noise over time." XXX} 
First, the volume flowrate is an integral quantity over the entire cross-section, which can flatten the fluctuations from the \HDS{}. Moreover, the spatial derivative on the flowrate is only first order in the governing equations of \LDS{} to benefit the noise smoothing. Coupled with fixed boundary values, the system can effectively eliminate noise over time.
Then, \HDS{}s advance in parallel starting from the same initial conditions with \LDS{} to compute their discrepancies at ending temporal points, which will be used as refinements in the following prediction step to correct \LDS{} (see Fig. \ref{fig:graphic} (b) and (c)). Notice that the iteration can be terminated early since we observed that the solution converges fast.

We customized \AlgAbb{} for the application of simulating zebrafish blood flow in this work. The present algorithm \ref{alg:zebra_MultiParareal} is essentially identical to the original \AlgAbb{} except one addition -- skip mapping if $Q^n_k -Q^n_{k-1} < \Delta_{\text{th}}$, where $\Delta_{\text{th}}$ a user-defined threshold. This modification is intended to hedge against the non-uniform particle distribution due to the explicit modeling of red blood cells (RBCs). Since the flowrate is computed using particle density, non-uniform particle distribution makes the flowrate slightly inaccurate, as demonstrated in the Y-bifurcation benchmark example. When the difference between two consecutive iterations $Q^n_k-Q^n_{k-1}$ is less than $10\%$, we found that skipping the mapping $\mathcal{R}\{Q^n_k\}$ avoids inaccurate re-initialization of temporal sub-domains. 

%%%%%%%%%%%%%%%%%%%%%%%%%%%%%%%%%%%%%%%%%%%%%%%%
\subsection{Multiscale framework} \label{sec:multiframework}

Coupling mesh-based/mesh-free models across scales can be challenging, since issues may arise from the innate stochasticity of meso-scale and the quantification of flow variable from the mesh-free model. In this section, we will explain in detail the coupling scheme between the two coupling models, namely DPD and 1D. For clarity, we will outline the interplay between the models in Section \ref{sec:coupling}. We describe the implemented software architecture that enables real-time data transfer in massive parallel simulations in the Appendix.

%%%%%%%%%%%%%%%%%%%%%%
\subsubsection{Velocity computation} \label{sec:fluxcomputation}

\begin{figure} 
	\centering
	\subfloat[]{\includegraphics[width=0.45\textwidth]{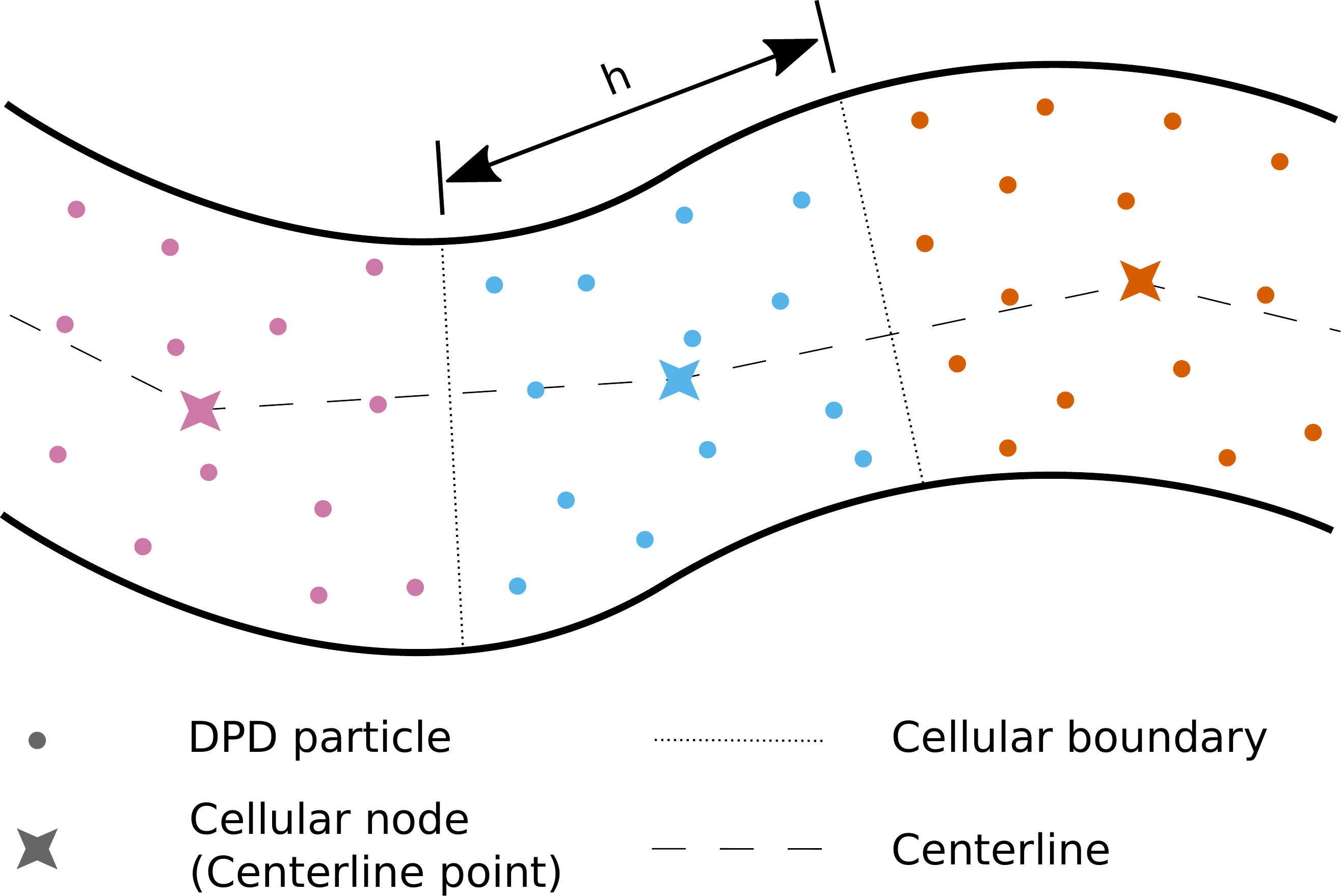}\label{one}} 
	\hspace{0.08\textwidth}
	\subfloat[]{\includegraphics[width=0.45\textwidth]{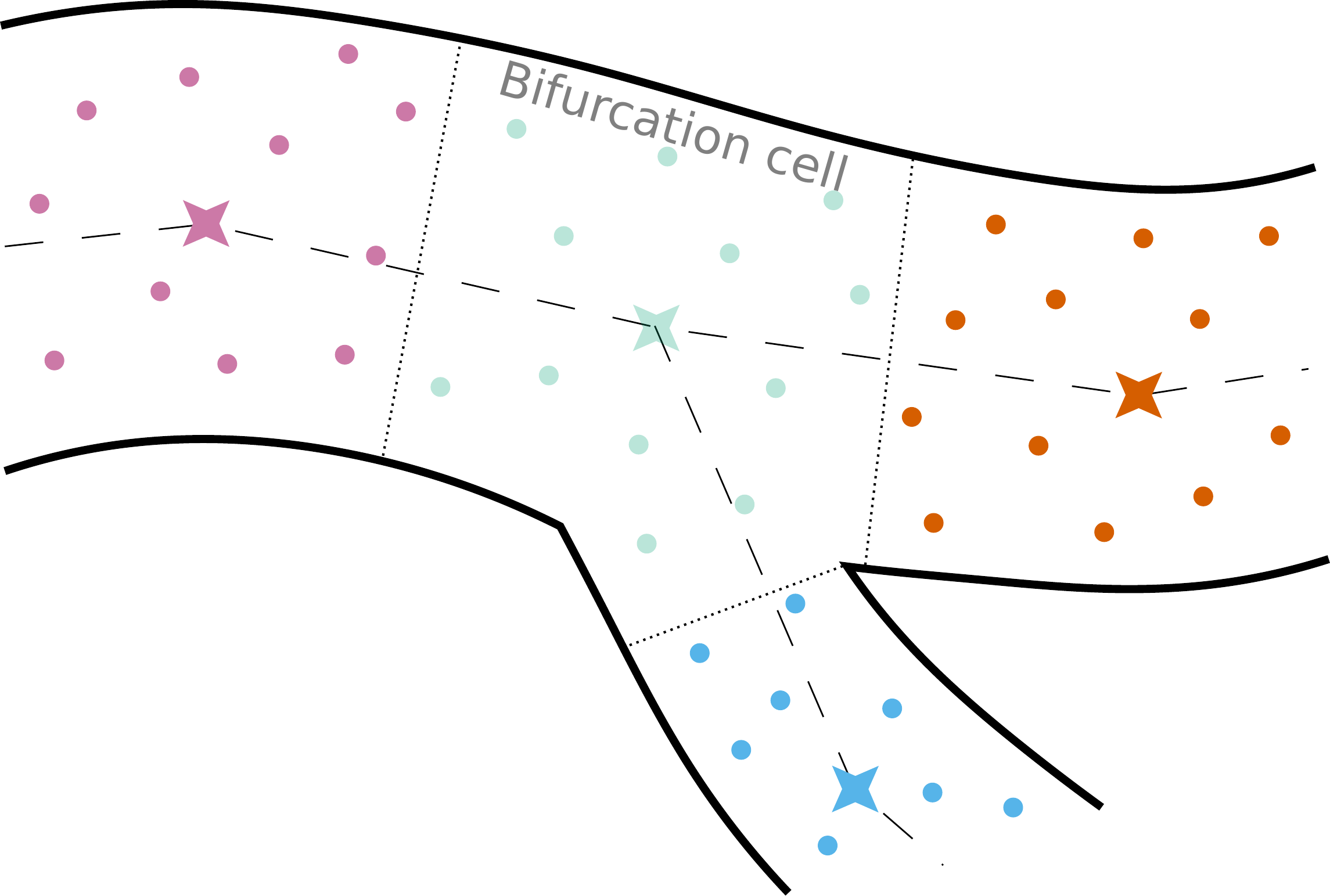}\label{bi}}
    \caption{\textbf{Particle-cell association.} (a) In \AlgAbb{}, the neighboring particles are lumped into one computational cell (referred as "cell"), and each cell is represented by its center referred to as the cellular node. The flowrate is computed on a cellular basis by averaging the dynamical state of particles the cell. (b) Because of the association scheme, the flowrate at a bifurcation cell is undefined and thus not computed. }
    \label{fig:lump}
\end{figure}

It is straightforward to compute the velocity in the 1D model. Since the computational domain of the 1D model is essentially a pipeline system, where each pipe is the centerline extracted from the 3D domain, the velocity can be readily computed at each centerline point as shown in Section \ref{sec:1Dmodel}. However, it takes additional steps to estimate the velocity for the DPD model. First, the 3D computational domain for the particle system can be decomposed into a list of non-overlapping computational cells (referred as ``cells" in the following context), where the neighboring particles are lumped into one cell. Each cell is represented by its center referred to as the cellular node, as illustrated in Figure \ref{fig:lump}. By averaging the dynamical state of particles in a cell as a whole, the mean flowrate for each cell can be approximated as 
\begin{equation} \label{eq:dpdmassflux}
Q^D \approx \sum_i \dfrac{\mathbf{v}_i \cdot \mathbf{\hat{n}}\cdot A}{{\rm Vol}\cdot\rho} = \sum_i \dfrac{\mathbf{v}_i \cdot \mathbf{\hat{n}}}{h\cdot\rho},
\end{equation}
where $\mathbf{v}_i$ is the velocity of particle $i$, $\mathbf{\hat{n}}$ is the direction of the centerline passing through the cell, and $A$ is the area of the cross-section. We refer to the cell thickness $h$ as the length of a computational cell along its centerline. The mean axial velocity is computed by $\sum{\mathbf{v_i}\cdot\hat{\mathbf{n}}}/({\rm Vol}\cdot\rho)$ with ${\rm Vol}=A\cdot h$ being the cell volume and $\rho$ being the number density of DPD particles. The mean flowrate is estimated by the product of the mean axial velocity and the area of cross-section $A$. Note that the velocity at bifurcation cells is not computed since they is undefined. Because of the correspondence of computational domains between the two models, we can establish connections, which enable multi-scale communications transported by the flow variable.

%%%%%%%%%%%%%%%%%%%%%%
\subsubsection{Coupling scheme} \label{sec:coupling}

\begin{table}     
  \centering
  \setlength{\tabcolsep}{10pt} % column spacing; Default value: 6pt
  \renewcommand{\arraystretch}{1.5} % row spacing; Default value: 1
	\begin{tabular} {c p{5cm} c p{8cm}} % textwrap
		\hline
		  & \textbf{1D model} & & \textbf{DPD model} \\
		\hline
		1 & & & Run the DPD simulation. \\
		2 & & & Down-scaling: Associate particles to their respective cells; compute flowrate $\{Q^\text{D}\}$ for each cell by summing over contributions from enclosed particles  \\
		3 & Fetch $\{Q^{D}\}$ on centerline points & $\blacktriangleleft$ & Push flowrate for each cell $\{Q^\text{D}\}$ on cellular nodes \\
		4 & Update $\{Q^{N}\}$ with corrections as described in \AlgAbb{} \\
		5 & Push $\{Q^{N}\}$ on centerline points from the next timestamp & $\blacktriangleright$ & Fetch updated flowrate $\{Q^\text{N}\}$ on cellular nodes \\
 		6 & & & Up-scaling: Tune particle velocity to match the corresponding cellular flowrate $\{Q^\text{N}\}$ using $\{Q^\text{D}\}$. \\
		  & Repeat from 1 & & Repeat from 1 \\
		\hline
	\end{tabular}
    \caption{\textbf{Outline of the coupling scheme between the 1D and DPD systems.} } 
    \label{tab:flow}
\end{table}

Outlined in Table \ref{tab:flow}, the coupling scheme starts with associating particles to their corresponding cells after a propagation in the DPD model. The flowrate ${Q^{D}}$ in DPD is then computed and pushed into the 1D model as the initial condition. Then, a series of 1D simulations using the fetched ${Q^{D}}$ as initial condition will be performed to compute an updated flowrate $Q^{N}$ with corrections in serial. Subsequently, the computed flowrate will be pushed back to the DPD model as the initial conditions, i.e., particle velocities, which is commonly referred to as ``up-scaling" in the context of multiscale coupling". In particular, a universal mapping cannot effectively match the target flowrates due to the stochastic nature of the particle system. Hence, we propose a modified mapping scheme pre-conditioned on the current state as following:
\begin{numcases}{\mathbf{v}^\text{New}_i \leftarrow} 
    \dfrac{Q^N}{A} \mathbf{\hat{n}} & , if $1-\alpha < F < 1 + \alpha$ \label{eq:method1} \\
    \mathbf{v}_i + (\mathbf{v}_i \cdot \mathbf{\hat{n}}) (\dfrac{Q^N}{Q^D}-1) \mathbf{\hat{n}} & , otherwise \label{eq:method2}, 
\end{numcases}
%{\color{red}XXX YH: IN THE CASE OF EQ(2), THE VELOCITY PROFILE IS NOT MAINTINED, BUT A UNIFORM VELOCITY IS IMPOSED ACROSS THE CROSS SECTION IN FS, AM I CORRECT ? XXX}
where $F \equiv (Q^N-Q^D)/Q^N$ and $\alpha$ is a pre-determined parameter taking a value between 0 and 1 to regulate the method selection. $A$ is the average area of a cell, which we approximate as the cross-sectional area at the cellular node.

The idea of this pre-conditioning mapping is to filter out numerical artifacts that may lead to divergence in this scheme. When the current flowrate $Q^D$ and the target flowrate $Q^N$ are on the same order of magnitude and non-zero, the particle velocity is modified according to Equation \eqref{eq:method2}. However, $Q^N/Q^D$ becomes unreasonably large if $Q^D$ is close to zero, where this scenario typically occurs in the first iteration when the particle system (or part of it) was under initialization. In this case, Equation \eqref{eq:method2} cannot accurately tune particle velocities. Instead, the particle velocity is modified according to Equ. \eqref{eq:method1} when $1-\alpha < F < 1 + \alpha$. For each combination of the signs of $Q^N$ and $Q^D$, we tabulate the preferred method given the range of $F$ in Table \ref{tab:methodselection}.

Since the 1D and the DPD model have their respective coordinate systems, a directional mapping $\mathbf{\hat{n}}$ for translating vectorial quantities between the solvers is necessary to couple them together. The flow orientation $\mathbf{\hat{n}}$ for each branch needs to be pre-assigned in the 1D model before performing simulations. However, the pre-assigned orientations will not affect the actual flow direction in the coupling scheme; the velocity value can be negative if there exists a mismatch. We illustrate more details on implementation in the Appendix.

\begin{table} 
    \centering
    \setlength{\tabcolsep}{14pt} % column spacing; Default value: 6pt
    \renewcommand{\arraystretch}{1.5} % row spacing; Default value: 1
    \begin{tabular}{c  c  c}
        \bf{Case} & \bf{Range} & \bf{Method} \\
        \hline
        \multirow{2}{*}{\makecell{$Q^N:-\longrightarrow+$ \\ $Q^D:-\longrightarrow+$}} &  
        $0<Q^D<\alpha Q^N \quad \leftrightarrow \quad 1-\alpha<F<1$ & 
        Equation \eqref{eq:method1} \\ 
        \cline{2-3}
         & $\alpha Q^N<Q^D  \quad \leftrightarrow \quad  F<1-\alpha$ & 
        Equation \eqref{eq:method2} \\ 
        \hline
        \multirow{2}{*}{\makecell{$Q^N:-\longrightarrow+$ \\ $Q^D:-\longleftarrow+$}} &  
        $-\alpha Q^N<Q^D<0 \quad  \leftrightarrow \quad  1<F<1+\alpha$ & 
        Equation \eqref{eq:method1} \\
        \cline{2-3}
         & $Q^D<-\alpha Q^N  \quad  \leftrightarrow  \quad 1+\alpha<F$ &
        Equation \eqref{eq:method2} \\
        \hline
        \multirow{2}{*}{\makecell{$Q^N:-\longleftarrow+$ \\ $Q^D:-\longleftarrow+$}} &  
        $\alpha Q^N<Q^D<0  \quad  \leftrightarrow  \quad  1-\alpha<F<1$ & 
        Equation \eqref{eq:method1} \\
        \cline{2-3}
         & $Q^D<\alpha Q^N  \quad  \leftrightarrow  \quad  F<1-\alpha$ &
        Equation \eqref{eq:method2} \\
        \hline
        \multirow{2}{*}{\makecell{$Q^N:-\longleftarrow+$ \\ $Q^D:-\longrightarrow+$}} &  
        $0<Q^D<-\alpha Q^N  \quad  \leftrightarrow  \quad  1<F<1+\alpha$ & 
        Equation \eqref{eq:method1} \\
        \cline{2-3}
         & $-\alpha Q^N<Q^D  \quad  \leftrightarrow  \quad  1+\alpha<F$ &
        Equation \eqref{eq:method2} \\
        \hline
    \end{tabular}
    \caption{\textbf{Mapping to the target flowrate in the up-scaling step is pre-conditioned on the current step.} We tabulated the preferred method for a given combination of $Q^N$ and $Q^D$. $\alpha$ is a pre-determined parameter between 0 and 1 regulating the method selection. $F$ is defined as $(Q^{N} - Q^{D})/Q^{N}$.}
    \label{tab:methodselection}
\end{table}

%%%%%%%%%%%%%%%%%%%%%%%%%%%%%%%%%%%%%%%%%

%%%%%%%%%%%%%%%%%%%%%%%%%%%%%%%%%%%%%%%%%%%%%%%%%%%%%%%%%%%%%%%%%%%%%%%%%%%%%%%%%%%%%%%%%%%%%%%%%%%%
\section{Results} \label{sec:numericalresults}

\subsection{Benchmark problem -- Y bifurcation} \label{sec:Ybifur}

\begin{figure} 
\centering
\includegraphics[width=0.5\textwidth]{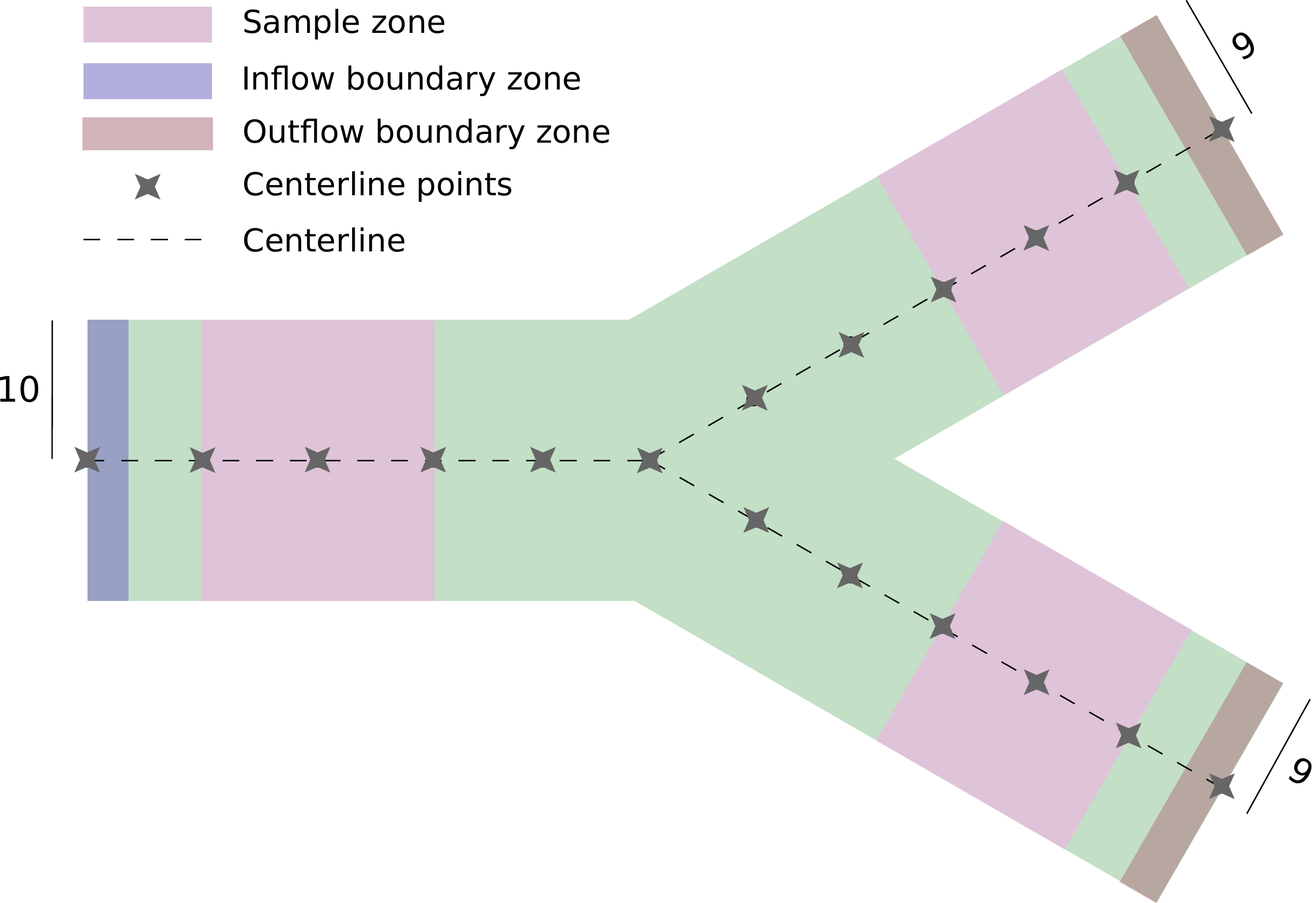}
\caption{\textbf{Central cut-plane of the 3-dimensional bifurcating channel.} The channel is composed of one parent pipe and two identical daughter pipes. For DPD, Dirichlet-type boundary conditions are enforced in the boundary zones. We sample statistics of the flow, such as velocity and flowrate, in sample zones.}
\label{fig:fiburcation}
\end{figure}

We first consider a Y-bifurcating channel composed of one parent pipe and two identical daughter pipes, with a branch half-angle $26$ degrees, as illustrated in Figure ~\ref{fig:fiburcation}. For DPD simulations, we impose Dirichlet boundary conditions at the inlet and outlets by prescribing velocity profiles in boundary zones. The velocity profiles and shear-stress are then calculated by sampling particles in the sample zones. The sample zones are placed away from the bifurcating point and boundaries so that the flow is fully developed in the sample zones and can be meaningfully compared. For 1D simulations, we can impose a Dirichlet boundary condition for velocity at the inlet, and impose the Windkessel boundary condition at the outlets with calibrated values. More details can be found in the Appendix. The results obtained from the proposed multiscale model are compared against the reference results, which are obtained from simulations without parallel-in-time. This allows us to directly assess the accuracy of our parallel-in-time framework. The deviation from the reference solution is quantified using the $l^2$ relative error $\epsilon$, expressed as
\begin{align} 
\epsilon &\equiv \frac{ \big\lVert Q^\text{ref} - Q^\text{sim} \big\rVert_{l^2} } {\big\lVert Q^\text{ref} \big\rVert_{l^2}} \equiv \frac{ \big( \sum_{i=1}^N (Q_i^\text{ref}-Q_i^\text{sim})^2 \big)^{1/2} } { \big( \sum_{i=1}^N (Q_i^\text{ref})^2 \big)^{1/2} } , \label{eq:relative_error_norm}
\end{align}
where $Q^\text{ref}$ and $Q^\text{sim}$ are the reference and simulated results, respectively. Moreover, we average the results from $40$ independent simulations to discount the thermal noise. The averaged results are then compared against those of the reference. Although the noise is still present post-averaging, it is reduced to a manageable magnitude so that the validity of \AlgAbb{} can be meaningfully inspected.

%%%%%%%%%%%%%%%%%%%%%%%%%%%%%%%%%%%%%%%%%%%%%%%%%%%%%%%%%%%%%%%%%%%%%%%

\subsubsection{Simple fluid} \label{sec:SimpleFluid}

\begin{figure} 
\centering
\includegraphics[width=0.5\textwidth]{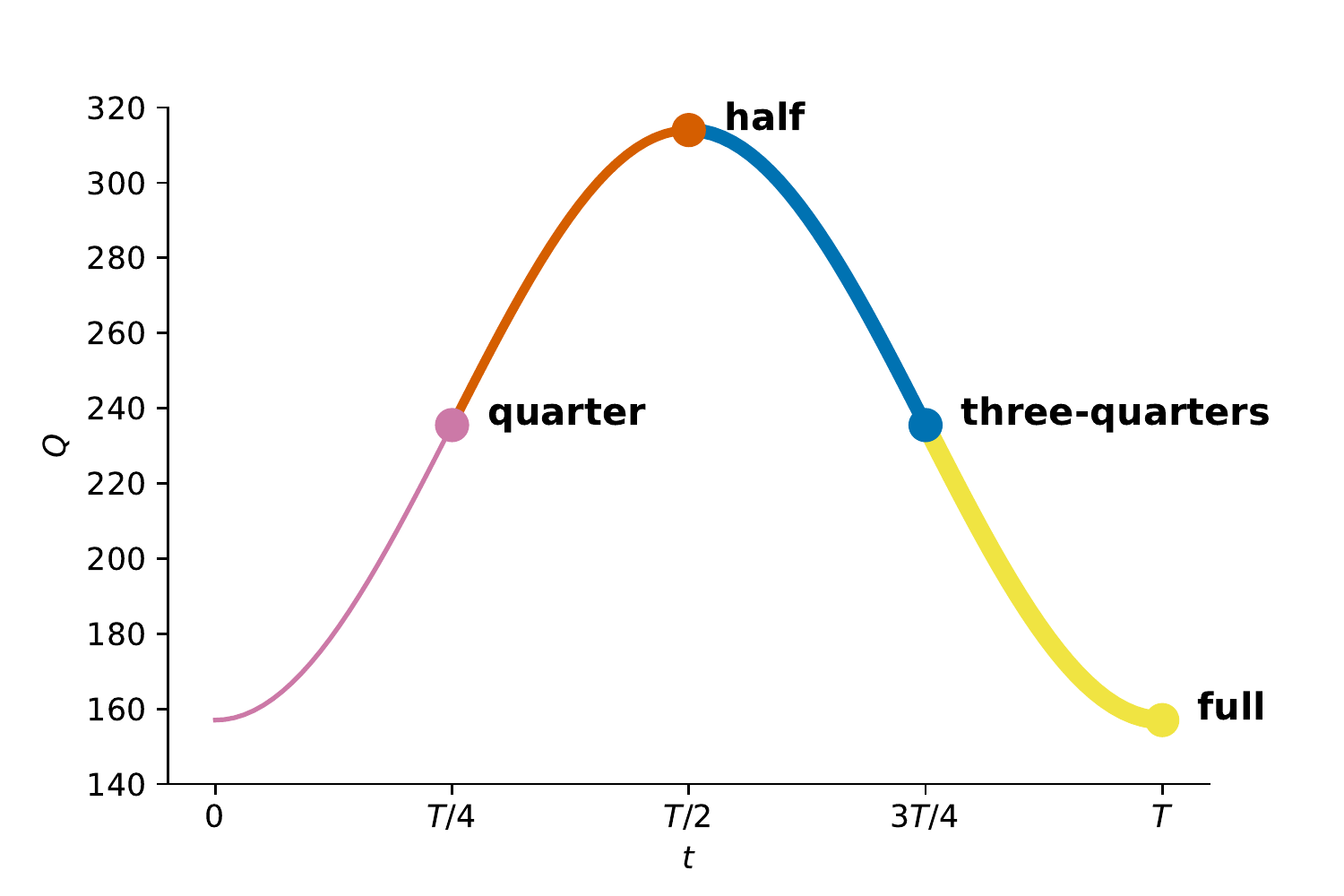}
\caption{\textbf{Time-dependent inlet boundary condition for the simple fluid example.} The flowrate ($Q$) at the boundaries are prescribed as a sinusoidal function of time. The peak $Q$ for the parent pipe is $314$ as shown here. Since each daughter pipes has a flowrate half of that of the parent pipe, its peak $Q$ is $175$. The temporal space, a full sinusoidal period, is divided into four quarters. The end of each quarter is marked on the plot. }
\label{fig:boundary_condition}
\end{figure}

We first consider a simple fluid flow in this Y bifurcation channel with Newtonian fluid assumption. The DPD pair interaction parameters are chosen as: $\alpha=30$, $\gamma=4.5$, $\sigma=3$, $s=2$, and $r_{c}=1$.  As shown in Fig. \ref{fig:boundary_condition}, we impose a Dirichlet boundary condition at the inlet for flowrate with a parabolic profile defined as $u(r)=c(R^2-r^2)$, where $r$ denotes the radius, and $c$ is the magnitude. In this example, $c_\text{inflow}$ and $c_\text{outflow}$ are $0.02$ and $0.0152$, respectively. For this simple fluid example, the flowrates at the boundaries are prescribed by adjusting the velocity of particles in the boundary zones to match the desired values. For 1D simulations, the inlet boundary condition is properly non-dimensionalized and imposed with density and dynamic viscosity of $1060$ $kg/m^{3}$ and $4$ $cP$.

%\color{red}{XXX YH: THE ASSUMPTION OF "A PARABOLIC PROFILE" IS JUSTIFIED WHEN THE TIME PERIOD OF THE PULSATION IS SUFFICIENTLY LONGER THAN THE TIME SCALE DETERMINED BY THE FLUID VISCOSITY AND THE PIPE RADIUS. I ASSUME THIS CONDITION HOLDS HERE, BUT NEED TO BE POINTED OUT.XXX} 
%\textcolor{blue}{The parabolic profile is just for boundary conditions, not the velocity profile in downstream.} 

\begin{figure} 
\centering
\includegraphics[width=\textwidth]{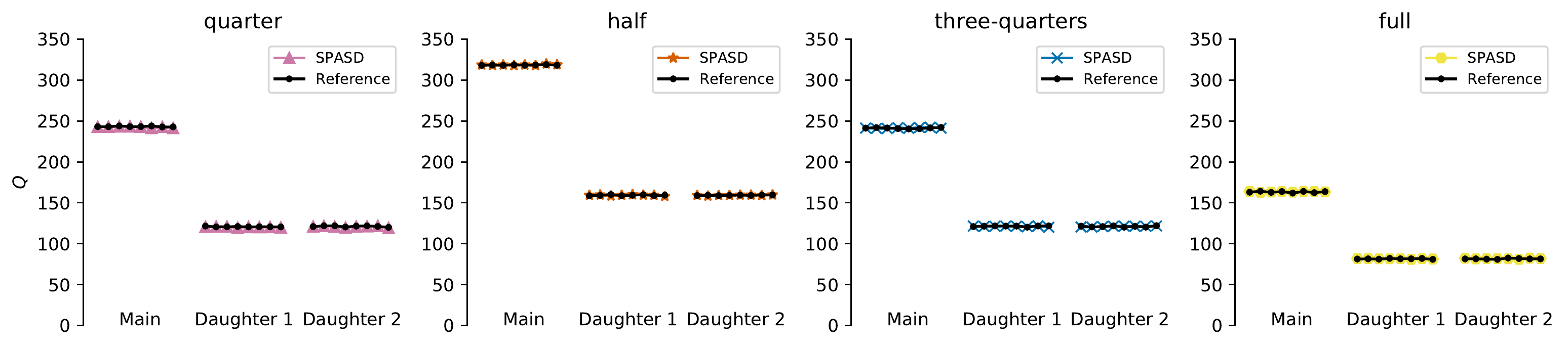}
\caption{\textbf{Flowrate at the end of each temporal sub-domain marked in Figure \ref{fig:fiburcation} in the simple fluid example.} The refined flowrate at iteration 3 in each pipe is compared against the reference flowrate obtained via serial simulations. The horizontal axis denotes the axial coordinate of sampled points on each branch. We see that \AlgAbb{} is able to produce solutions very close to the reference.}
\label{fig:SimpleFluidMassFlux}
\end{figure}

In Fig. \ref{fig:boundary_condition}, the temporal domain is divided into four non-overlapping sub-domains each of which is associated with an independent DPD system running concurrently. That is, the systems are initialized at $t=\{0,\;T/4,\;T/2,\;3T/4\}$, and ended at $t=\{T/4,\;T/2,\;3T/4,\;T\}$, respectively. We plot the refined flowrate at iteration 3 from DPD simulations in Figure \ref{fig:SimpleFluidMassFlux} at the end of each temporal sub-domain. The horizontal axis denotes the axial coordinate of sampled points on each branch. Furthermore, we tabulate the $l^2$ relative errors $\epsilon$ in Table \ref{tab:simple_fluid_error} and found that $\epsilon$ is less than $1\%$ for all sub-domains. Additionally, the numerical convergence is shown by the errors are on the same order across iterations. All of these results indicate a good agreement between the reference results and that of \AlgAbb{}.

\begin{table}
\centering
\begin{tabular}{c  c | c | c | c | c}
 & & \bf{quarter} & \bf{half} & \bf{three-quarters} & \bf{full} \\
\hline
\multirow{3}{*}{$\epsilon$} & Iteration 1 & $0.50\%$ & $0.51\%$ & $0.70\%$ & $0.70\%$ \\
 & Iteration 2 & $0.58\%$ & $0.37\%$ & $0.40\%$ & $0.75\%$ \\
 & Iteration 3 & $0.55\%$ & $0.36\%$ & $0.44\%$ & $1.01\%$ \\
\hline
\end{tabular}
\caption{\textbf{$l^2$ relative error of the flowrate defined in Equation \eqref{eq:relative_error_norm}, for the simple fluid example.} $\epsilon$ less than $1\%$ for all sub-domains indicates good agreement between the reference and \AlgAbb{}. }
\label{tab:simple_fluid_error}
\end{table}

\begin{figure} 
\centering
\subfloat[]{\includegraphics[width=1.0\textwidth]{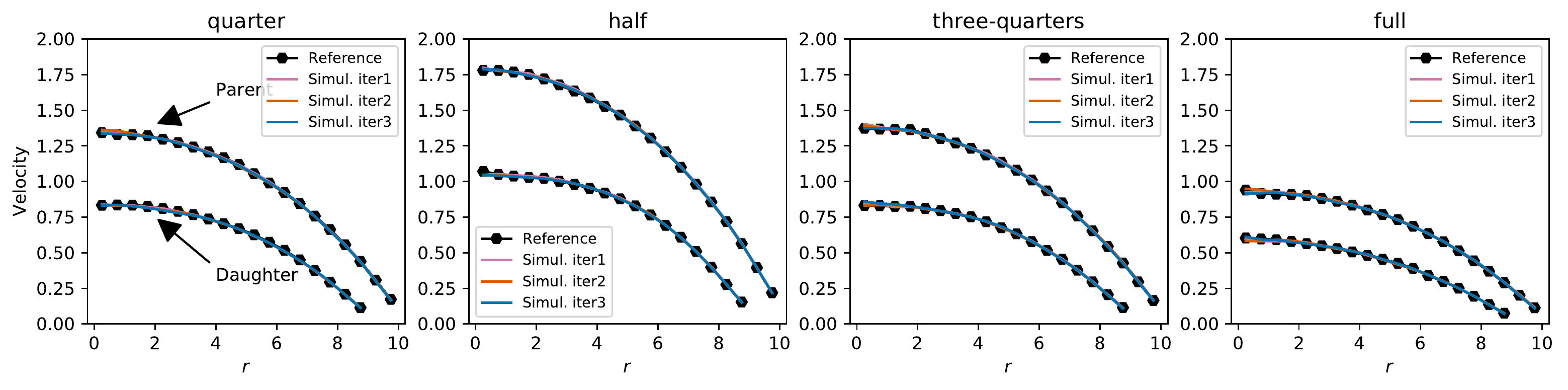}\label{fig:Vprof}} \\
\subfloat[]{\includegraphics[width=1.0\textwidth]{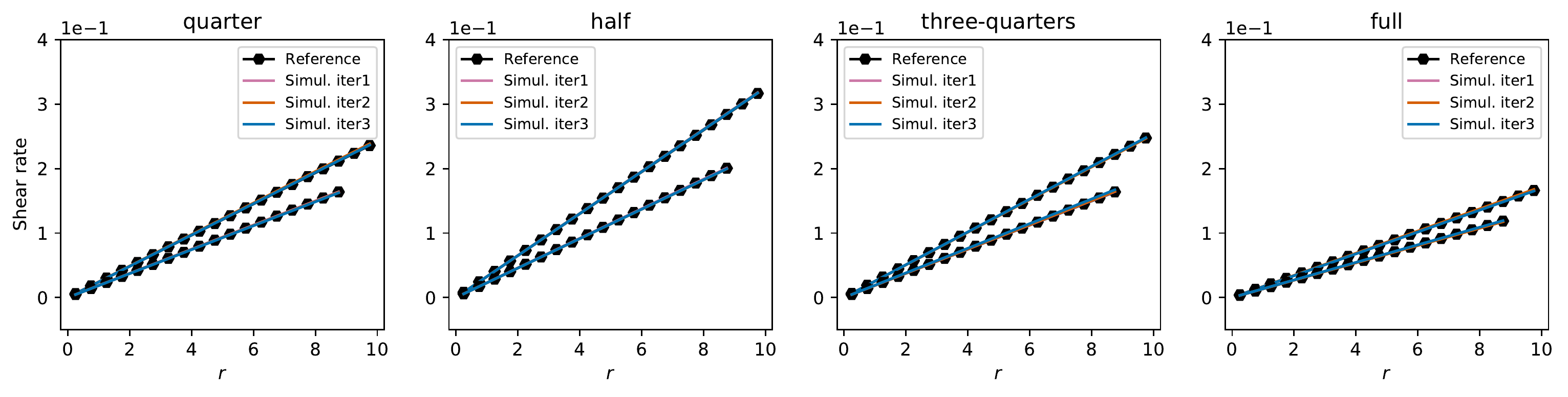}\label{fig:Shear}}
\caption{\textbf{(a) Velocity profiles and (b) shear rates from the DPD system at specific times marked in the bifurcation channel Figure \ref{fig:fiburcation}.} We plot the results from three iterations to show convergence, and see that the flow from \AlgAbb{} is very close to that from the reference simulation.}
\label{fig:SimpleVprof}
\end{figure}

Moreover, we compare the velocity profiles and shear rates from DPD in Figure \ref{fig:SimpleVprof} for further verification. A shear rate are obtained by first fitting a velocity profile to a parabolic equation, and then calculating its analytic gradient. Again, we found the reference and \AlgAbb{} solutions match well for all iterations. Importantly, we observe that the results and errors stay approximately the same regardless of the number of iterations, indicating that one iteration is sufficient for convergence.

%%%%%%%%%%%%%%%%%%%%%%%%%%%%%%%%%%%%%%%%%%%%%%%%%%%%%%%%%%%%%%%%%%%%%%

\subsubsection{Complex fluid: Blood flow} \label{sec:ComplexFluid}

\begin{table}
\centering
\begin{tabular}{c | c | c | c | c | c | c}
 & & $\alpha$ & $\gamma$ & $\sigma$ & $s$ & $r_\text{c}$ \\
\hline
Wall & Any & 60 & 45 & 2.916 & 2 & 1 \\
Solvent & Solvent & 60 & 45 & 2.916 & 2 & 1 \\
Solvent & RBC & 4 & 45 & 2.916 & 2 & 1 \\
RBC & RBC & 4 & 30 & 2.381 & 0.5 & 1.5 \\
\hline
\end{tabular}
\caption{\textbf{DPD parameters of pair interaction in Y-bifurcation example.}}
\label{tab:ComplexPairCoeff}
\end{table}

In this example, we test our proposed model on the same Y-bifurcating channel with complex fluids where the fluid is a mixture of single particles and discrete RBCs at a hematocrit of $60\%$. Illustrated similarly in Figure \ref{fig:boundary_condition}, the complex fluid flow is driven by prescribing a sinusoidal flowrate over time at boundaries. The maximum flowrate at the parent and each daughter boundary is approximately $63$ and $31.5$, respectively. We impose Dirichlet boundary conditions with an exponential velocity profile $u(r)=ae^{br}+c$, where $a$ and $c$ are adjusted to match the magnitude of the desired flowrate. The DPD parameters are set as shown in Table \ref{tab:ComplexPairCoeff}. The boundary conditions for 1D simulations are imposed in a similar way with the simple fluid example. Noticing that the 1D model only serves as a low-fidelity guide where certain levels of simplifications can be tolerated, the parameter setting in this example is inherent from the simple fluid case (Newtonian fluid), where a laminar blood flow with with parabolic velocity profile is assumed to be valid. In particular, the parabolic parameter $\alpha=\frac{4}{3}$.

\begin{figure} 
\centering
\includegraphics[width=\textwidth]{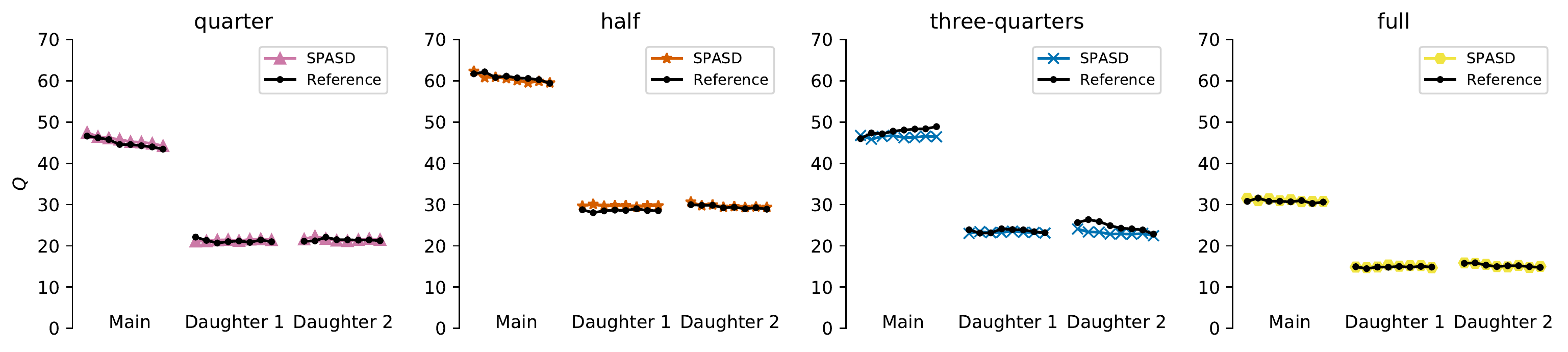}
\caption{\textbf{Flowrate at the end of each temporal sub-domain marked in Figure \ref{fig:boundary_condition}, for the blood flow example.} This plot corresponds to iteration 3. The horizontal axis denotes the axial coordinate of sampled points on each branch. Due to the uniformity of particle distribution in the complex fluids, the flowrate and velocity profiles match those of the reference less closely. We will discuss and expand on this topic further in Section \ref{sec:discussion}.
}
\label{fig:ComplexMassFlux}
\end{figure}

\begin{figure}
\centering
\includegraphics[width=1.0\textwidth]{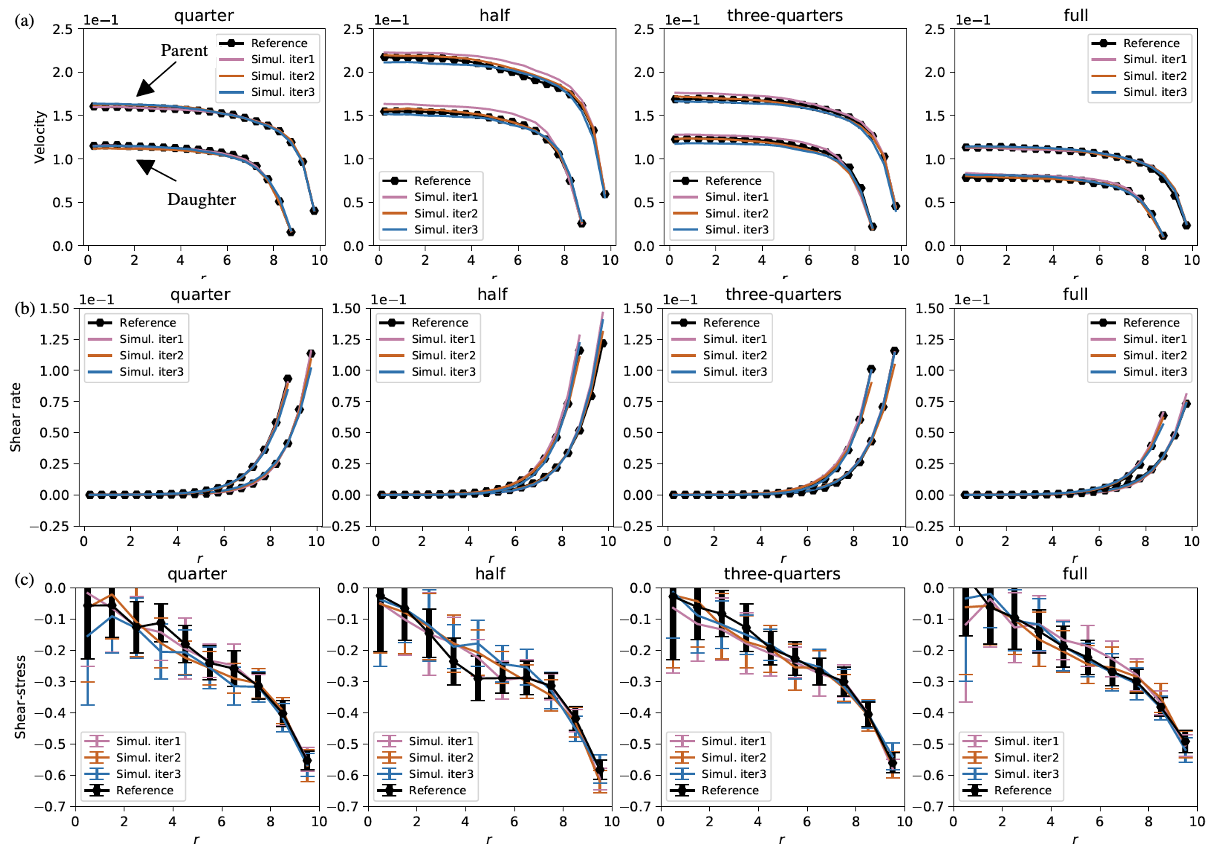}
\caption{\textbf{(a) Velocity profiles and (b) shear rates and (c) shear stress at specific times marked in Figure \ref{fig:boundary_condition} for the blood flow example.} We plot these quantities and compare them to those of the reference simulations. Shear stress is measured from $50$ independent runs, and the error bar denotes one standard deviation from the mean.}
\label{fig:ComplexFluid}
\end{figure}

\begin{table}
\centering
\begin{tabular}{c  c | c | c | c | c}
 & & \bf{Quarter} & \bf{Half} & \bf{Three-quarters} & \bf{Full} \\
\hline
\multirow{3}{*}{$\epsilon$} & Iteration 1 & $0.93\%$ & $5.34\%$ & $3.77\%$ & $2.35\%$ \\
 & Iteration 2 & $2.60\%$ & $2.18\%$ & $3.14\%$ & $2.61\%$ \\
 & Iteration 3 & $2.01\%$ & $1.99\%$ & $4.25\%$ & $1.58\%$ \\
\hline
\end{tabular}
\caption{\textbf{$l^2$ relative error of flowrate for the blood flow example.}}
\label{tab:Complex_fluid_error}
\end{table}

We compare the refined flowrate for complex fluids at different snapshots in Fig. \ref{fig:ComplexMassFlux}. This plot corresponds to iteration 3. The horizontal axis denotes the axial coordinate of sampled points on each branch. The tabulated $l^2$ relative error $\epsilon$ is shown in Table \ref{tab:Complex_fluid_error}. In general, the results from the reference and \AlgAbb{} correspond well. However, due to the non-uniformity of particle distribution in the complex fluids, the flowrate and velocity profiles match those of the reference less precisely than that in the simple fluid case. We will discuss and expand on this topic further in Section \ref{sec:discussion}. Furthermore, we present the velocity profiles and shear rates with the reference values against radius $r$ at different sites and times in \ref{fig:ComplexFluid}(a) and (b). Since the shear stress at the wall is of particular interest owing to its mechanobiological effects on  signaling activation, we compute the shear stress $\tau$ by averaging over bins that decompose the 3D domain. Also, we calculated the mean and variance of shear stresses from $50$ independent simulations and plotted the results in Fig. \ref{fig:ComplexFluid}(c). As $r$ increases, $\tau_\text{reference}$ matches $\tau_\text{SPASD}$ much better. It is noticeable that the shear stress close to the center has a larger variance caused by the fact that the concentrations of RBCs and solvent particles are higher towards the center.

%\begin{figure} 
%\centering
%\includegraphics[width=0.5\textwidth]{Plots/Flux_time_dependent_complex_fluid_illustration.pdf}
%\caption{Time-dependent boundary condition for the complex fluid example. Like the simple fluid example, the flowratees at the boundaries are prescribed as a sinusoidal function of time. Each of the daughter pipes has a flux half of that of the parent pipe, which is shown on the y-axis. The temporal space of a period is divided into four quarters, with the end of each quarter marked on the plot.}
%\label{fig:cfbc}
%\end{figure}

%%%%%%%%%%%%%%%%%%%%%%%%%%%%%%%%%%%%%%%%%%%%%%%%%%%%%%%%%%%%%%%%%%%%%%%%%%%%%%%%%%%%%%%%%%%%%%%%%%%%

\subsection{Realistic vascularture -- zebrafish hindbrain} \label{sec:zebrafishhindbrain}

\begin{figure}
    \centering
	\includegraphics[width=0.8\textwidth]{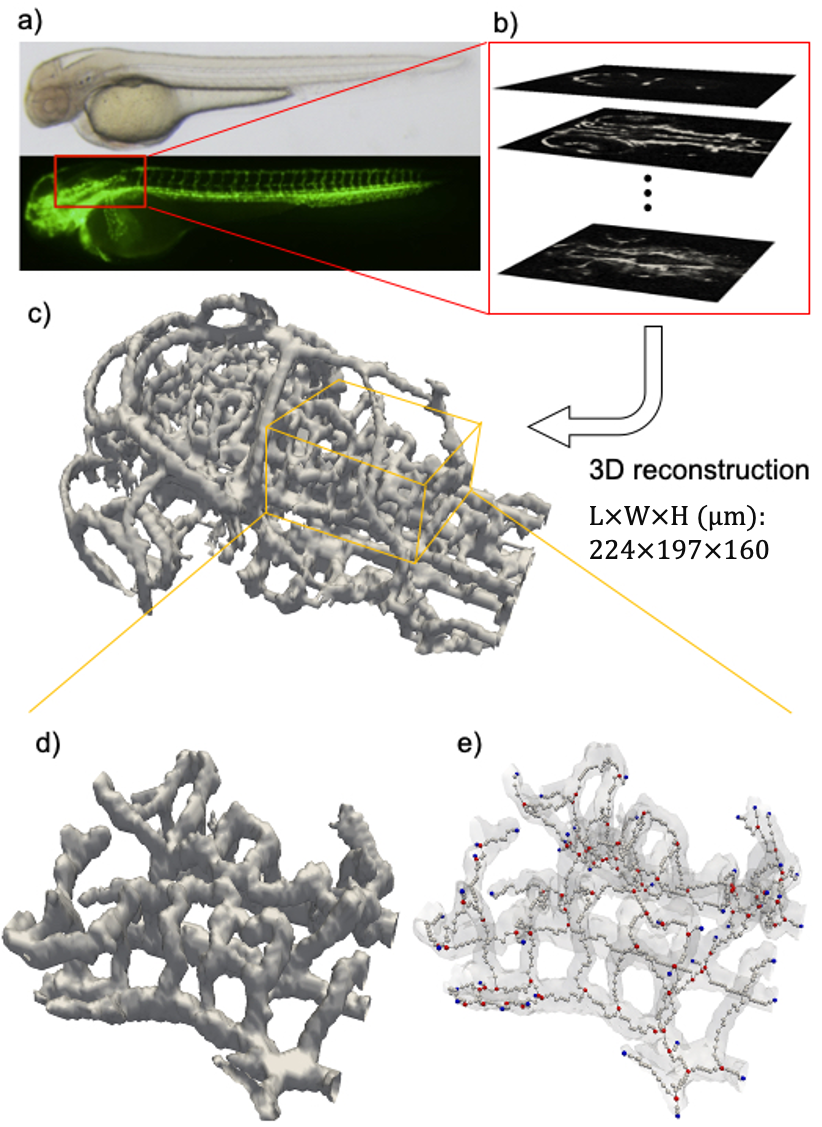} 
    \caption{\textbf{3D geometry reconstruction of zebrafish brain vasculature. }
    (a) vascular network in the entire body (b) a stack of 2D confocal microscope images of brain vasculature (c) reconstructed 3D vascular network in the brain (d) hindbrain vasculature for DPD simulation (e) extraction of the center line of each microvessel for 1D analysis (blue dot: end point, red dot: branching point, white dot: vessel center).
%    Geometry reconstruction for the high- and low-dimensional representations. We choose a zebrafish larva at approximately two days postfertilization (dpf) as our model. Specifically, we focus on the hindbrain of a living zebrafish. 
%    The images were obtained with confocal microscopy by our collaborators at the National Cerebral and Cardiovascular Center in Japan. We convert the scan into a digital replica, which is then used to construct a particle system that is identical to the real vasculature. Additionally, the centerline of the these vessels is extracted for the continuum solver.
    }
    \label{fig:leading}
\end{figure}

\begin{figure} 
\centering
\includegraphics[width=0.9\textwidth]{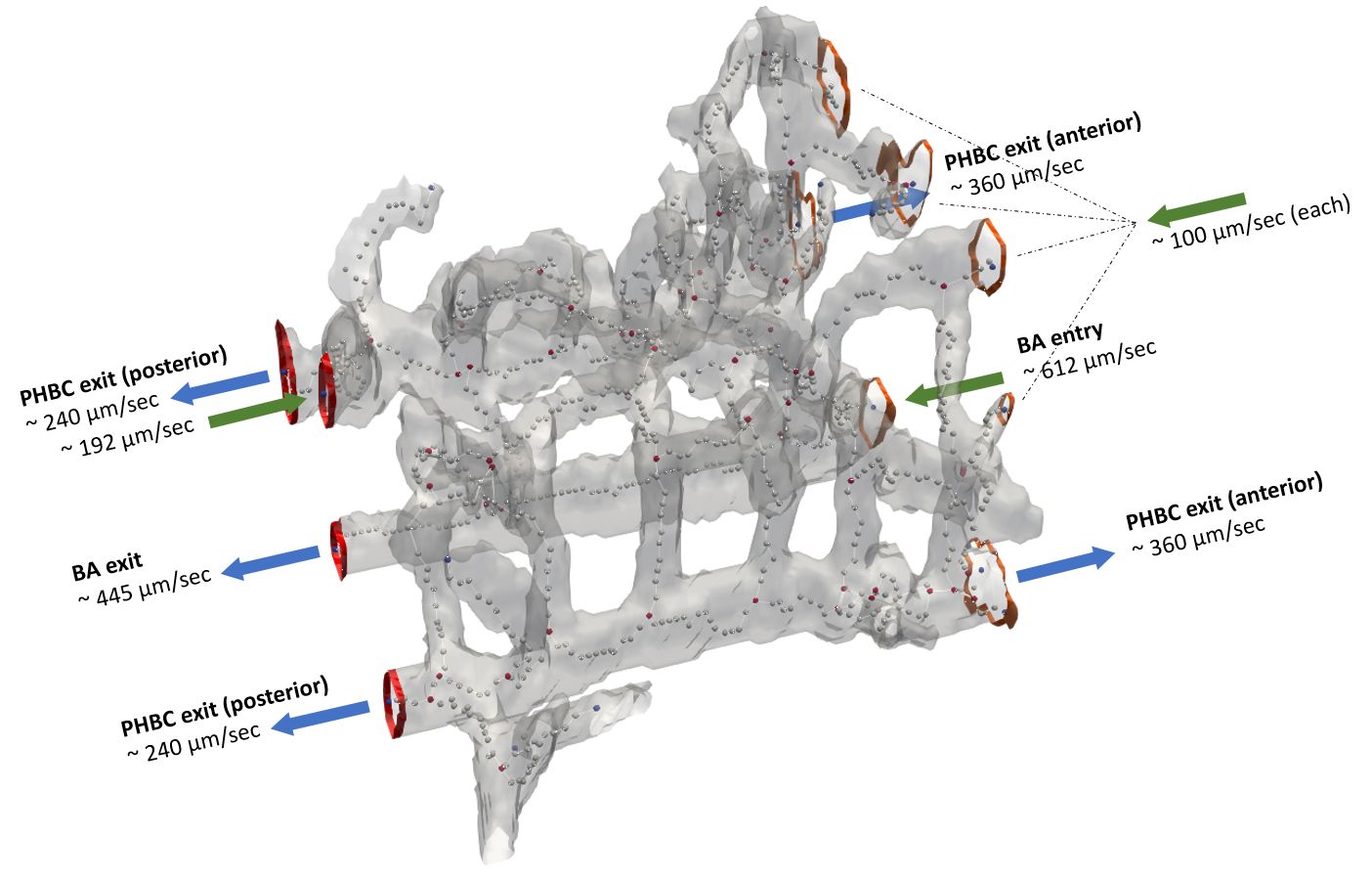}
\caption{\textbf{Overlap of the realistic 3D geometry and the 1D centerline}. The blue nodes represent boundaries of the volume, and the red nodes represent bifurcations. The boundaries are highlighted in red, and the flow direction is marked with arrows. To mimic the pulsatile flow from heartbeats, the flowrates at inlets and outlets are prescribed as periodic time-dependent functions shown in Figure \ref{fig:heartbeat_BC}. }
\label{fig:3d_1d_overlap}
\end{figure}

In this section, we consider a zebrafish hindbrain as the biological model (Fig.~\ref{fig:leading}) to test our method on a complex domain. To visualize the zebrafish brain vasculature, we use a transgenic (Tg) zebrafish line, Tg(kdrl:mCherry-CAAX)ncv8, that can label plasma membrane of endothelial cells (ECs) specifically by expressesing mCherry-CAAX under the control of kdrl (flk1) promoter~\cite{Wakayama}. For 3D confocal imaging, the zebrafish embryos are mounted in 1\% low-melting agarose poured on a 35-mm-diameter glass-base dish (Asahi Techno Glass). Confocal images are taken with a FluoView FV1000 confocal upright microscope system (Olympus) equipped with water-immersion XLUMPlan FL N 20x/1.00 NA objective lens (Olympus) regulated with FluoView ASW software (Olympus). In order to visualize the behavior of red blood cells (RBCs) flowing inside the vessels, we use double-transgenic Tg(kdrl:GFP);Tg(gata1:DsRed) embryos at 55 hours post-fertilization (hpf) that can label ECs and RBC with GFP and DsRed, respectively. High-speed movie images of zebrafish blood flow are taken at 68 frames per second (fps) using light sheet microscopy. Light sheet imaging is performed with a Lightsheet Z.1 system (Carl Zeiss) equipped with a water immersion 20x detection objective lens (W Plan Apochromat, NA 1.0), dual sided 10x illumination objective lenses (LSMF, NA 0.2), a pco.edge scientific CMOS camera (PCO) and ZEN software. We track RBCs manually with ImageJ software (National Institute of Health) in order to measure the velocity of RBCs. 14-26 RBCs (from 3 embryos) are tracked so as to obtain the blood flow velocities at the boundaries of the current computational domain. Blood velocity and direction at some boundaries are measured $\textit{in vivo}$ over a period of time (see Fig.~\ref{fig:3d_1d_overlap}). These measurement data are used for the boundary conditions in the present simulations.

As shown in Fig.~\ref{fig:leading}, the three primary vessels, running along the body, carry blood from the heart to the rest of the body. The secondary vessels, sitting on top the primary vessels, branch from the primaries and circulate blood circumferentially. Our image analysis reveals 95 distinctive branches with 57 bifurcating junctions in our region of interest. At 2 days post-fertilization (dpf) developmental stage, the vessel diameters range from 10 to 20 micrometers, and the region is merely 224x197x160 micrometers in size (see Fig.~\ref{fig:3d_1d_overlap}). The compactness of the region and the complexity of the vascular network make it challenging for the 1D solver to model it accurately, but it is an ideal benchmark for our method.

\subsubsection{Computational Setup} \label{sec:zebrafishsetup}

\begin{table}
\centering
\begin{tabular}{c | c | c | c | c | c}
 & $\alpha$ & $\gamma$ & $\sigma$ & $s$ & $r_\text{c}$ \\
\hline
Wall \& Any & 20 & 45 & 2.916 & 2 & 1.5 \\
Solvent \& Solvent & 20 & 45 & 2.916 & 2 & 1.5 \\
Solvent \& RBC & 4 & 45 & 2.916 & 2 & 1.5 \\
RBC \& RBC & 4 & 30 & 2.381 & 0.5 & 1.5 \\
\hline
\end{tabular}
\caption{\textbf{DPD parameters of pair interaction in zebrafish hindbrain example.}}
\label{tab:zebrafishPairCoeff}
\end{table}

Our reconstructed 3D particle system contains 17,244,362 particles in total -- 5,989,445 fixed particles as walls, 9,477,881 solvent particles and 3,554 RBCs, each of which is made of 500 particles. The hematocrit of our particles system matches the true hematocrit of 2 dpf larva at $43\%$ \cite{Lee2017}. To form a closed system with periodic boundary conditions, the inlets and outlets along the spinal direction are connected to a reservoir. As the outlets push blood into the reservoirs, the inlets take the blood from the reservoirs to replenish. To keep the implementation simple, we impose a no-flux boundary condition on the other four sides (i.e., top, bottom, front and back). A DPD boundary method for arbitrarily complex geometries developed by Li et al.~\cite{2018Li_ADPD} is used to automatically handle the 3D zebrafish brain vasculature. Table~\ref{tab:zebrafishPairCoeff} shows the values of DPD parameters. The centerline for 1D is extracted based on the reconstructed 3D geometry and is used as the 1D computational domain. Since the 1D model only provides a coarse approximation of the blood flow, we set 1D parameters to the same values as the complex fluid case from above.

\begin{figure} 
\centering
\includegraphics[width=0.5\textwidth]{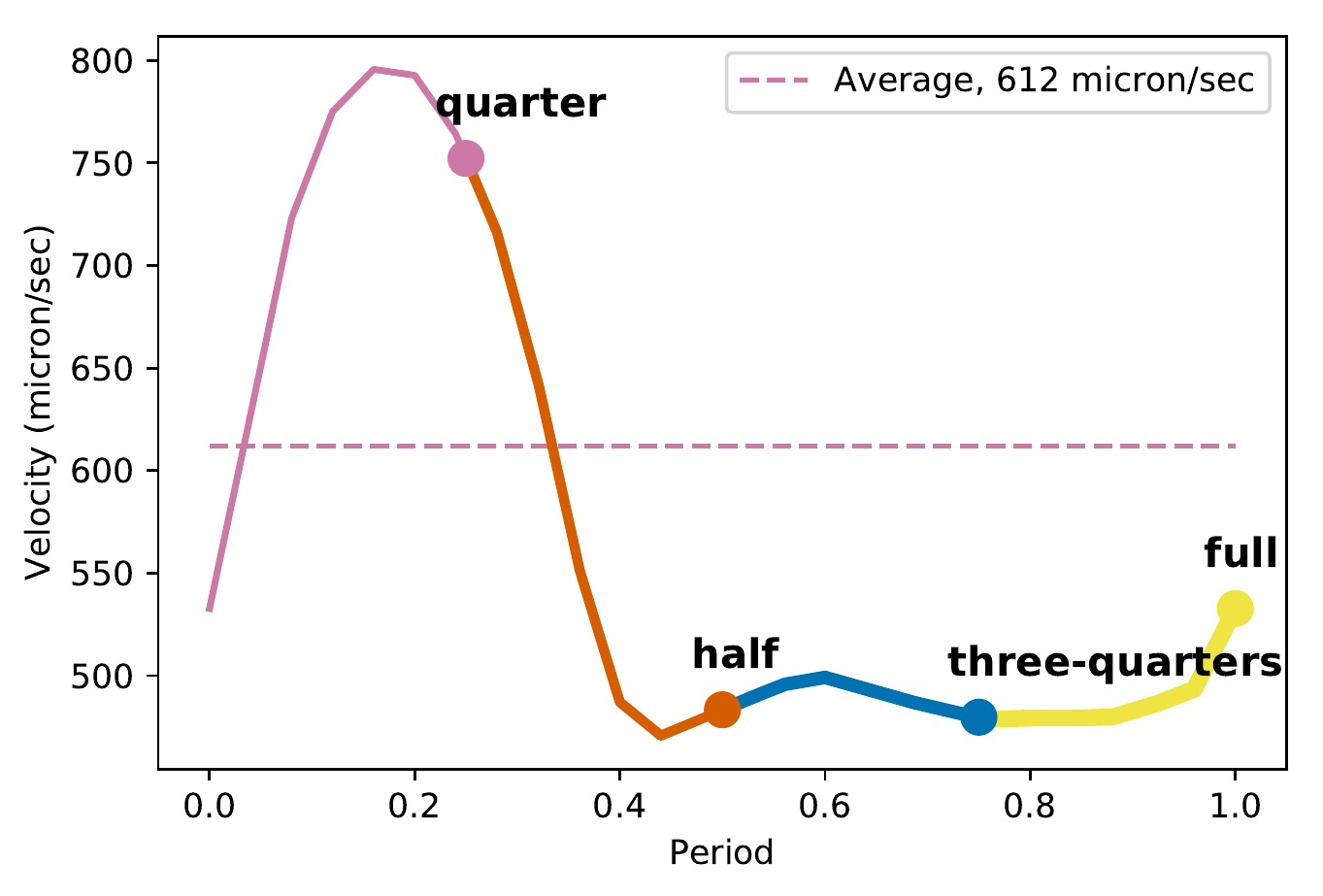}
\caption{\textbf{A standard pulsatile flowrate in a single period.} We normalize the standard pulsatile flow in accordance with the mean and maximum values of blood velocity at the boundaries. %For example, an inlet with an average speed of 612 $\mu$m/sec has a maximum velocity of 795.6 $\mu$m/sec (peak) and a minimum of 470.7 $\mu$m/sec (trough). The temporal space of a full cycle is divided into four quarters, with the end of each quarter marked on the plot.
}
\label{fig:heartbeat_BC}
\end{figure}

We prescribe the same Dirichlet boundary conditions for both models, namely pulsatile flows mimicking the realistic conditions as shown in Figure \ref{fig:heartbeat_BC}. The zebrafish was at 2 dpf, and the heart rate was approximately 175 beats per minute \cite{Kopp2005}. We re-normalized the flowrate profiles for each outlet/inlet so that the mean and max value match the experimental measurements. We further assume that the pulsatile flow is in-phase at inlets and outlets, although in reality there may exist a small phase difference.

To avoid numerical overflow and boost computational accuracy, the 1D and DPD system use their respective reduced units, which can be mapped to physical units. For the particle solver, the length scale is chosen to be $[L^{\text{DPD}}] = 1 \times 10^{-6}$m. This means that one unit length in the DPD system is equivalent to $10^{-6}$ meters of physical length. By matching the viscosity of real blood to that of the DPD fluid, the time scale is calculated to be $[T^{\text{DPD}}]=1.2225 \times 10^{-3}$ s. On the continuum solver side, the length and time scales are chosen so that the flowrate matches that of the real blood in zebrafish. In this example, we used $[L^\text{1D}]=1 \times 10^{-2}$m and $[T^\text{1D}]=1$s for stability and accuracy. The details on mapping to physical units can be found in Appendix \ref{sec:AppendixB}.

\begin{figure} 
\centering
\includegraphics[width=0.8\textwidth]{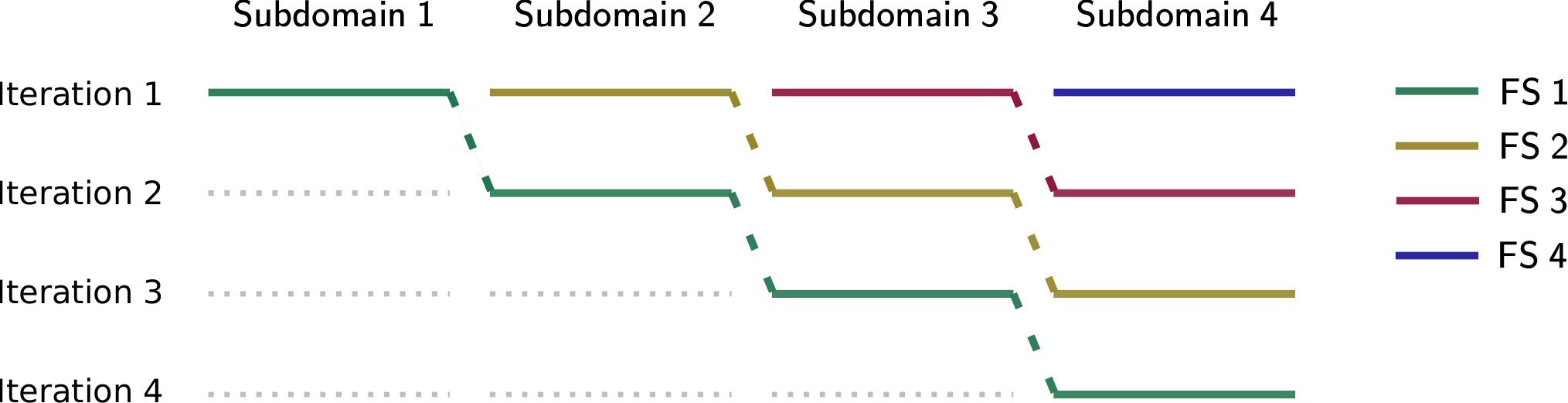}
\caption{\textbf{The temporal domain is divided into four sub-domains.} For the first iteration, \textit{FS 1, 2, 3} and \textit{4} propagates the systems concurrently, one for each sub-domain. At the end of the first iteration, the solution in sub-domain 1 is accurate and does not need to be iteratively corrected. If a second iteration is needed, \textit{FS 1} moves down to sub-domain 2. The same applies to \textit{FS 2} and \textit{3}. The convergence in the global domain is guaranteed within four iterations.}
\label{fig:shiftup}
\end{figure}

We divide the temporal domain into four sub-domains, as illustrated in Figure \ref{fig:shiftup}. For the first iteration, \textit{FS 1, 2, 3} and \textit{4} propagates the systems concurrently, one for each sub-domain. If the computational results in sub-domain 1 is accurate at the end of first iteration, it is unnecessary to proceed to further iterations. If further iterations are needed, \textit{FS 1} moves down to sub-domain 2 and the same applies to \textit{FS 2} and \textit{3}. The convergence in the global domain is guaranteed within four iterations.

%%%%%%%%%%%%%%%%%%%%%%%%%%%%%%%%%%%%%%%%%%%%%%%%%%%%5
\subsubsection{Validation} \label{sec:zebrafishval}

\begin{figure} 
\centering
\includegraphics[width=0.8\textwidth]{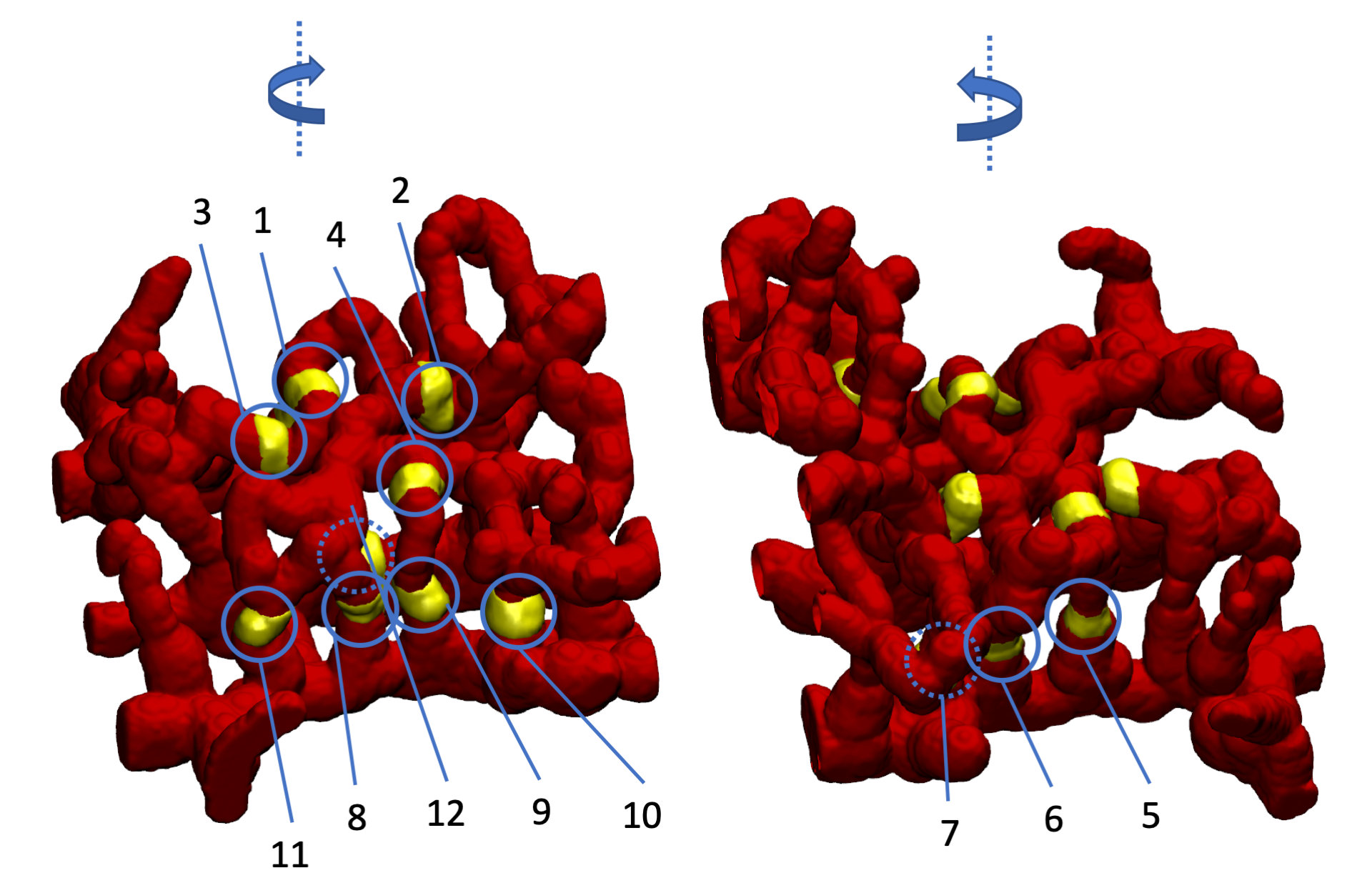}
\caption{\textbf{12 representative sites are marked in yellow.} Averaged velocity in \AlgAbb{} at these sites are extracted and compared against the reference results.}
\label{fig:specific_sites}
\end{figure}

\begin{figure}
    \centering
	\includegraphics[width=0.5\textwidth]{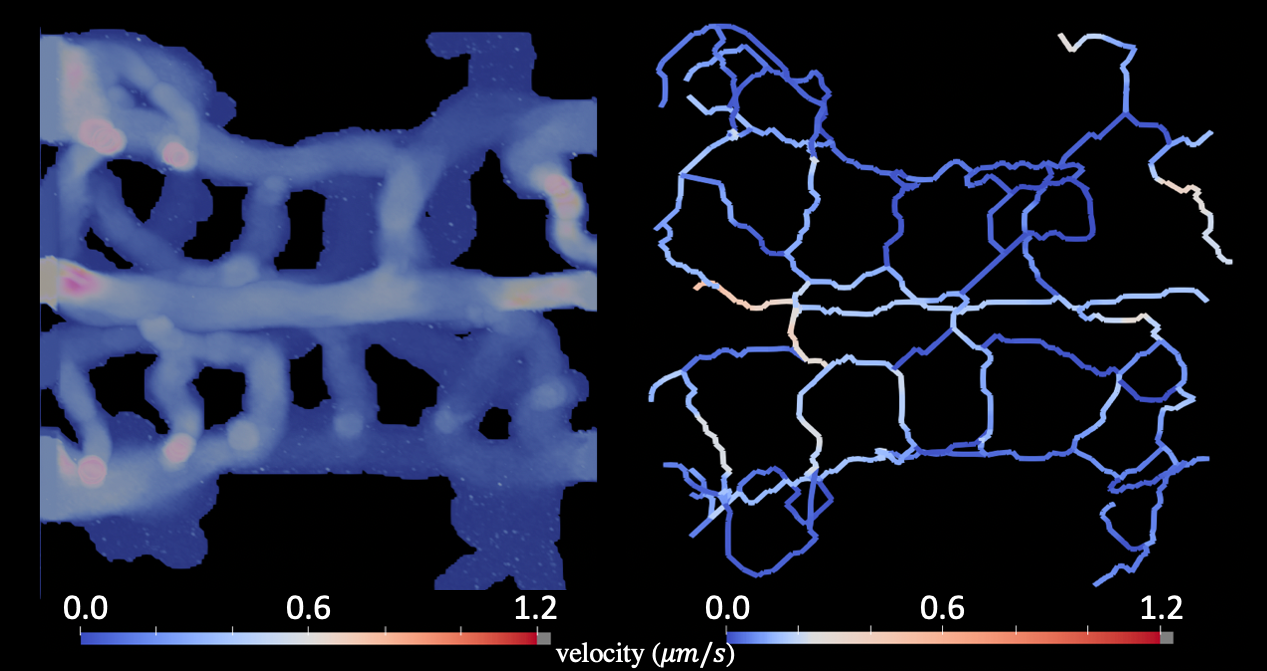} 
    \caption{\textbf{Qualitative comparison for the velocity field from the reference (left) and a standalone 1D simulation with the same boundary conditions.} The reference is obtained from an expensive DPD simulation without parallel-in-time. }
    \label{fig:velo_comp}
\end{figure}

\begin{figure} 
\centering
\includegraphics[width=0.8\textwidth]{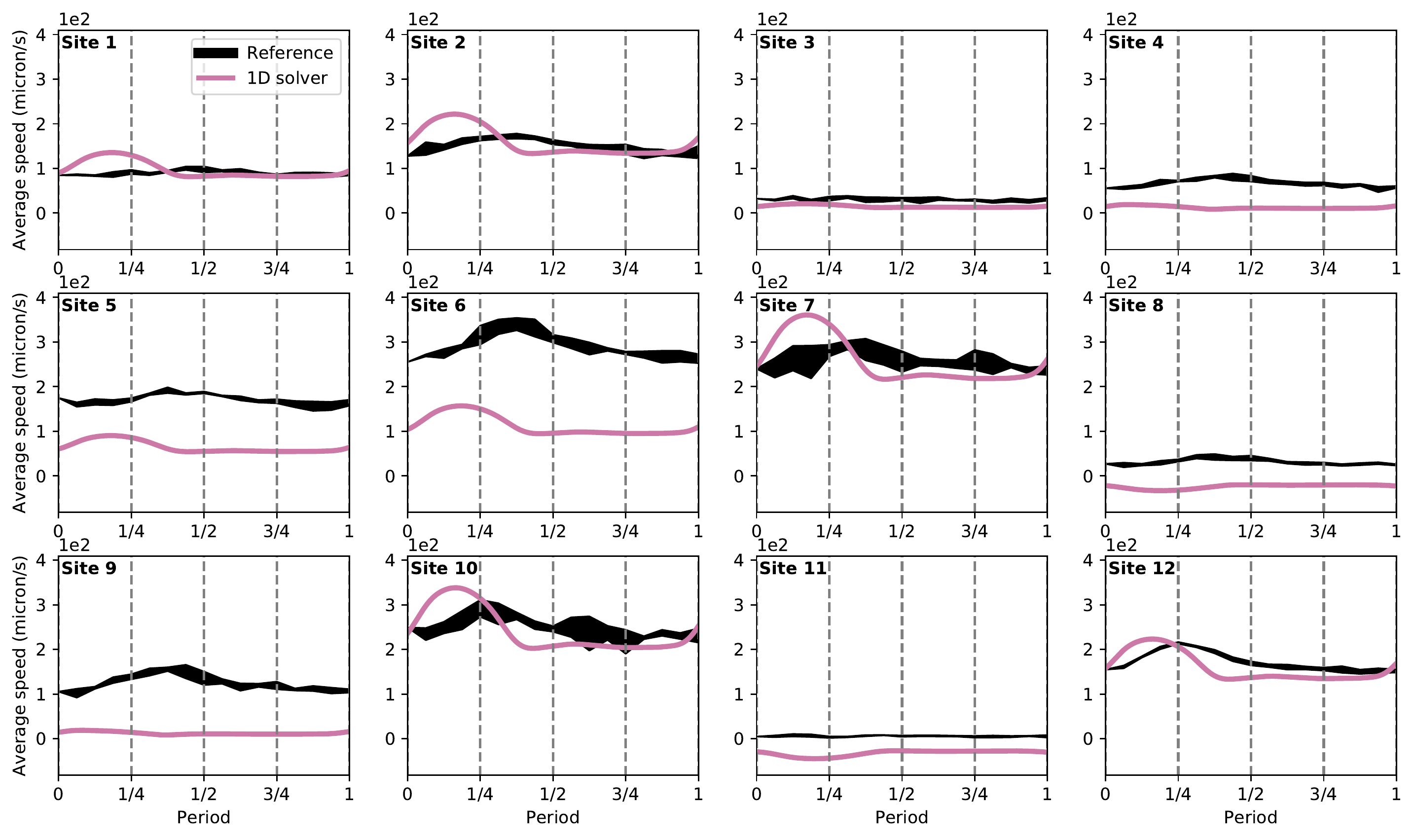}
\caption{\textbf{Quantitative comparison of predicted velocity from CS (labeled as `\textit{1D solver}') and FS (labeled as `\textit{Reference}') at the 12 sites.}}
\label{fig:fine_VS_coarse}
\end{figure}

We first simulate a full heart period using the 1D and DPD models independently. A qualitative comparison of the velocity field is presented in Fig. \ref{fig:velo_comp}. Overall, we can observe an agreement between the (a) DPD and (b) 1D model both with light color at the top left and top right sites (inlets and outlets), indicating a faster blood flow. To be more quantitative, the computed velocities are then compared at 12 representative sites shown in Fig. \ref{fig:specific_sites}, which include sites in the upstream and downstream. Fig. \ref{fig:fine_VS_coarse} presents a comparison of computed velocity from 1D and DPD where the results from DPD are averaged over 3 repetitive simulations. The extreme values at each timestep are recorded and plotted in Fig. \ref{fig:fine_VS_coarse} as well. The results from DPD simulations are considered as the reference solution, which fluctuate mildly due to the non-uniformity in the fluid flow. In general, the flow in 1D model has very little phase difference: the systoles for all vessels are approximately synchronized with little phase shift and with the inflow boundaries (Fig. \ref{fig:heartbeat_BC}). However, in the reference solution, there exists a noticeable phase differences at those sites, which may be due to the assumptions (e.g., neglecting branching angles) and simplifications of blood composition in the 1D model.

\begin{figure} 
\centering
\includegraphics[width=0.8\textwidth]{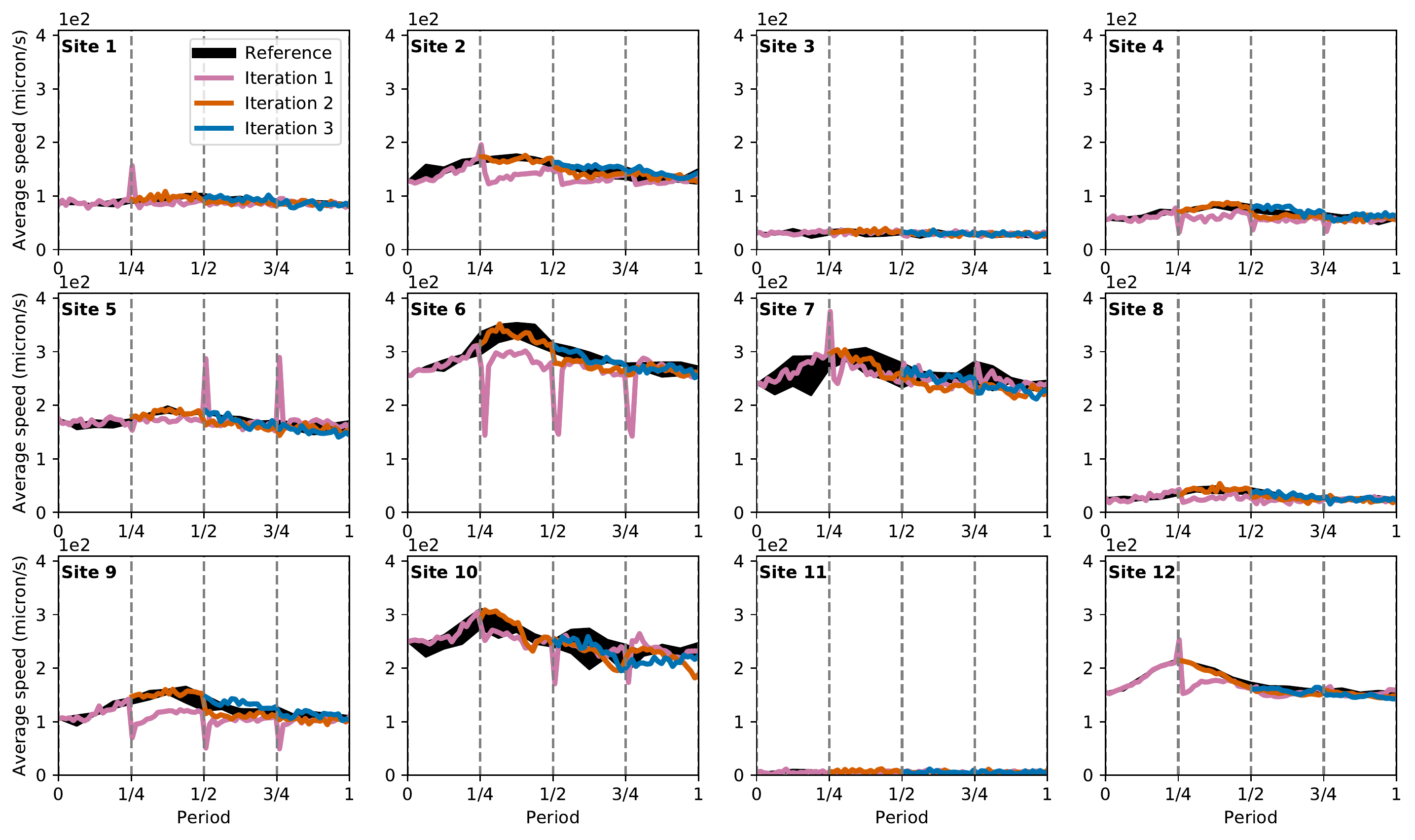}
\caption{\textbf{Comparison of averaged velocity in \AlgAbb{} at the 12 sites and the reference data.} The resulting average speed at 12 sites marked in Figure \ref{fig:specific_sites} are shown here. The reference data comes from three standalone FS simulations for comparison.}
\label{fig:Ref_VS_SPASD}
\end{figure}

\begin{figure} 
\centering
\includegraphics[width=0.8\textwidth]{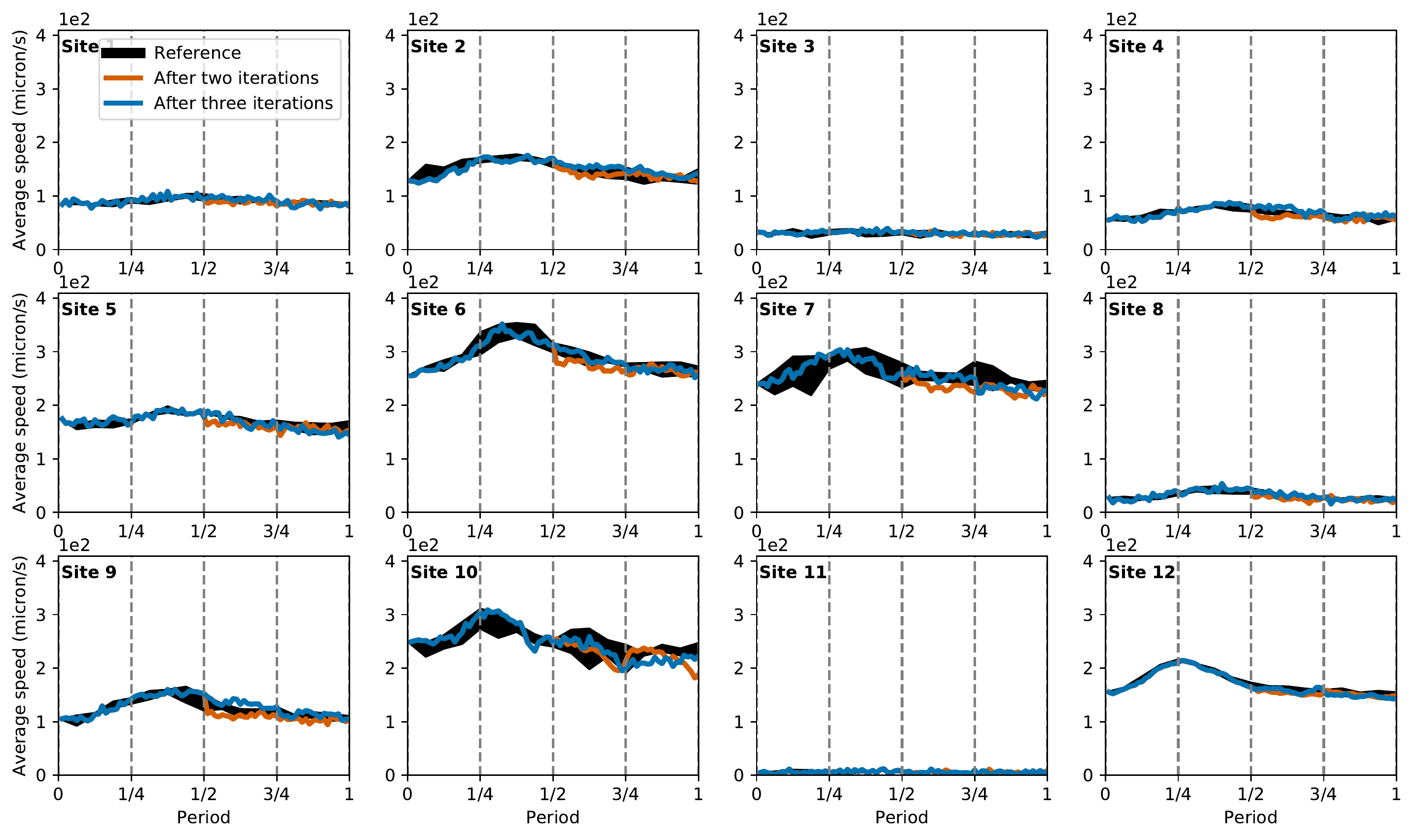}
\caption{\textbf{Results in Figure \ref{fig:Ref_VS_SPASD} in the second and third iterations are re-plotted.} The results are very close to the reference solutions after only two iterations. We run a third iteration for demonstration purposes.}
\label{fig:Ref_VS_SAPSD_STI}
\end{figure}

%Here, we employ spatial decomposition (SD) in conjunction with \AlgAbb{} for further acceleration. 

We then simulate a full heart period with \AlgAbb{}, and compare the values of averaged velocity from DPD simulations in the temporal sub-domains against the reference values at those sites. Since we divide the temporal domain into four sub-domains, \AlgAbb{} solution must converge in less than four iterations to be beneficial. Therefore, we run three iterations and plot the results against the reference solutions in Figure \ref{fig:Ref_VS_SPASD}. The results exhibit strong oscillations in velocity at interfaces of sub-domains in iteration 1, suggesting gaps between the two models. Subsequently, blood flow velocity in the network then moves back to the reference as it recovers from the inaccurate initial value. Notice that the speed of recovery is dependent on factors such as the amount of deviation from the reference and its magnitude. However, it is not a linear response by the fact that a small local perturbation can have cascading reactions across the network and cause a lengthy equilibrium recovery time in complex networks. \textit{Iteration 2} is supposed to be initialized with values calculated via \AlgAbb{}. Because the new values are within $10\%$ threshold from that of last iteration, no re-initializations were carried out at those sites. Instead, DPD models simply continue to march in time for the corresponding sub-domains. \textit{Iteration 2} skips the first sub-domain since the solution is obtained via FS alone. It continues from where \textit{Iteration 1} stops in \textit{sub-domain 1}. The same goes for the other sub-domains. After only two iterations, the result is very close to the reference solution. To demonstrate  convergence, we plot the reference solution and the results after two iterations in Fig. \ref{fig:Ref_VS_SAPSD_STI}. We run a third iteration and plot the results on the same figure for demonstration purposes. The difference after two and three iterations is less pronounced, meaning that \AlgAbb{} can be terminated after the second iteration.

% \begin{table}
% \centering
% \begin{tabular}{c | c | c | c | c | c }
%  & \AlgAbb{} WT (s) & $T_{\text{1D}}$, 1D WT (s) & $T_{\text{DPD}}$, DPD WT (s) & Speedup & Speedup Bound \\
% \hline
% % Y-bif iter. 1 (Simple) & 4 & 4 & 271 & 2 & $\sim$4 & 4\\
% % Y-bif iter. 1 (Complex) & 4 & 4 & 1789 & 2 & $\sim$4 & 4 \\
% Zebrafish iter. 1 & 2190 & 114 & 2076 & 3.79 & 4\\
% Zebrafish iter. 2 & 2153 & 86 & 2067 & 1.91 & 2 \\
% % Zebrafish iter. 3 & 2135 & 57 & 2078 & 1.28 & 1.33 \\
% % Zebrafish iter. 4 & 2103 & 29 & 2074 & 0.97 & 1 \\
% \hline
% \end{tabular}
% \caption{\textcolor{red}{\textbf{Speedup factor for the zebrafish simulations}. We decompose the temporal domain (TF) into four sub-domains where each domain are performed on 3 GPUs. Wall times (WT) of 1D, DPD, and \AlgAbb{} are also provided. The speedup is bounded by the number of sub-domains and  computed by discarding the time taken by the 1D solver.}}
% \label{tab:zebrafishSpeedup}
% \end{table}

\begin{table}
\centering
\begin{tabular}{c | c | c | c }
Zebrafish & \AlgAbb{} WT (s) & Extra Speedup & Extra Speedup Bound\\
\hline
iter. 1 & 2190 & 3.79 & 4\\
iter. 2 & 2153 & 1.91 & 2\\
\hline
\end{tabular}
\caption{\textbf{Extra speedup factor for the zebrafish simulations}. We decompose the temporal domain into four subdomains where each subdomain is performed on three NVIDIA Tesla V100 GPUs. The wall time (WT) of \AlgAbb{} and theoretical extra speedup bound is also provided with the measured speedup. The same simulation with only the DPD solver running in serial ($T_{\text{SD}}^{\text{serial}}$) would take 8301 seconds. }
\label{tab:zebrafishSpeedup}
\end{table}

\textbf{Remark:} Given a fixed-size problem, the parallel speedup using conventional spatial decomposition (SD) is limited by Amdahl’s law~\cite{al-hayanni2020amdahl}, meaning that we cannot gain any speedup by simply using more computational resources. This work aims to use the temporal decomposition (\AlgAbb{}) in conjunction with SD to achieve extra speedup when more computational resources become available. We define the extra speedup factor of \AlgAbb{} as
\begin{equation} \label{eq:speedup}
S = \frac{T_{ \text{SD} }^{\text{serial} } }{T^{\AlgAbb{}}},
\end{equation}
in which $T^{\AlgAbb{}}$ is the wall time of \AlgAbb{}, and $T_{\text{SD}}^{\text{serial}}$ is the wall time of SD being serial in time.  
%that the same simulation with only the DPD solver running in serial would take. 
For a meaningful comparison, the number of GPUs to clock $T_{\text{SD}}^{\text{serial}}$ is the same as that of each temporal subdomain in \AlgAbb{}. The speedup factor can be written as
\begin{equation} \label{eq:speedup_theo}
S = \frac{T_{ \text{SD} }^{\text{serial} } }{T^{\AlgAbb{}}} = \frac{\sum\limits_{i=1}^{N} \tau_{i}^{\text{DPD}}} { \sum\limits_{j=1}^{K}(\tau_{j}^{DPD} + \tau_{j}^{1D} + \tau_{j}^{\text{Op}})} \lessapprox \frac{N}{K},
\end{equation}
where $\tau_{i}^{\text{DPD}}$, $\tau_{i}^{\text{1D}}$, and $\tau_{i}^{\text{Op}}$ are the wall time taken by the DPD, 1D solver, and operations (noise-filtering, up-scaling, and down-scaling) at iteration $i$. $N$ is the number of temporal sub-domains and $K$ denotes the number of iterations to convergence. Ideally, $\tau_{i}^{\text{DPD}}$ keeps the same at different iteration, and the total wall time per iteration is disproportionately dominated by $\tau_{}^{\text{DPD}}$. Hence, the theoretical extra speedup is bounded by $\frac{N}{K}$.

For the zebrafish problem, we tabulated the wall time of \AlgAbb{}, measured extra speedup, and theoretical extra speedup in Table~\ref{tab:zebrafishSpeedup}. This particular benchmark was run on 12 NVIDIA Tesla V100 GPUs, where each temporal domain was performed on three GPUs. The ``serial'' benchmark takes 8301 seconds on three GPUs. As shown in the table, the measured extra speedup reaches at 3.79 and 1.91 in the first and second iterations, while the theoretical bound is 4 and 2, respectively. The measured extra speedup is very close to the theoretical bound, indicating that $\tau_{i}^{\text{1D}}$ is minuscule compared to $\tau_{i}^{\text{DPD}}$ for each iteration. Since the error between the second and third iterations is less pronounced, we terminated the simulation after two iterations.

%%%%%%%%%%%%%%%%%%%%%%%%%%%%%%%%%%%%%%%%%%%%%%%%%%%%%%%%%%%%%%%%%%%%%%%%%%%%%%%%%%%%%%%%%%%%%%%%%%%%
\section{Summary \& Discussion} \label{sec:discussion}

In this paper, we demonstrated how to accelerate massive Lagrangian simulations of blood flow in a zebrafish hindbrain by employing the parallel-in-time algorithm \AlgAbb{}. The framework was first demonstrated on a simple Y-bifurcation geometry. Physical quantities, i.e., velocity, shear rate, shear stress and flowrate, were computed and compared against the reference solutions. Specifically, the $l^2$ relative error on flowrate is less than $1\%$ for simple fluid and less than $3\%$ for complex fluid on average. We then demonstrated our framework on a reconstructed zebrafish hindbrain, which is a complex vascular network with 95 branches and 57 bifurcations. Comparatively, \AlgAbb{} is better suited for simulating of simple fluids than for complex fluids. As demonstrated in Y-bifurcation in Section~\ref{sec:Ybifur}, the flowrate and velocity profiles for a simple fluid match closer to those of the reference, which is largely due to the uniformity of particle distribution. In contrast, RBCs are modeled explicitly in the complex fluid. Particle non-uniformity is thus more pronounced because the local particle density is increased by the presence of the cells. In addition, the non-uniform scattering of cells causes non-uniform particle distribution at the network scale, which enlarges the discrepancy between flowrate from \AlgAbb{} and the reference in Figure~\ref{fig:ComplexMassFlux}. Despite this, the maximum normalized $l^2$ error is only on the order of one percent in our complex fluid example, and it does not grow over iterations. The above results demonstrate the efficacy and efficiency of our proposed multiscale coupling scheme \AlgAbb{} in accelerating a Lagrangian simulation of blood flow informed with \textit{in-vivo} imaging on a 2 dpf zebrafish, which provides a time-budgeting way to investigate mesoscopic biological processes with simulations.

%However, there are certain limitations for \AlgAbb{}. 
When simulating a large system with limited sources, SD offers better efficiency than \AlgAbb{}. This does not contradict the limitations of spatial decomposition listed in Section \ref{sec:intro}. Instead, this is drawn from our past benchmark on large-scale particle simulations, where we have demonstrated that SD scales very well with limited computational resources due to the absence of the CS and the need for a second iteration. However, when ample computational resources are available, the combination of \AlgAbb{} and SD offers better efficiency than SD alone for long-time simulations. Moreover, we still need to further investigate the optimal way to balance spatial and temporal decomposition. That is, for a fixed system with limited resources, we should maximize the overall efficiency by allocating the number of spatial and temporal subdomains. Also, the overall efficiency depends on other factors, such as machine architecture, GPU/CPU models, the efficiency of the computer program, etc. A further systematic study on maximizing the efficiency of \AlgAbb{} is worth pursuing.

%(XXX Keep it here or Results? XXX)
For the zebrafish example, we note that the extra speedup is defined as the computational gain by employing temporal decomposition. Three GPUs per temporal domain is a good balance between available resources and computational efficiency. Admittedly, the zebrafish example serves only as a proof of concept to primarily show how \AlgAbb{} can further accelerate a long-time mesoscopic simulation in a complex system together with conventional spatial decomposition. As shown in~\cite{Blumers2019}, the inter-nodal communication time will eventually nullify the benefits of additional cores for a large dynamical system with long-time integration. However, \AlgAbb{} does not suffer from such issues, which makes it a more scalable algorithm for long-time simulations. A higher speedup can be achieved if more computational resources are available so that the long temporal domain can be divided into more subdomains. For example, if one is interested in studying the hemodynamics of zebrafish in 100 cycles, the extra speedup can grow linearly with the increase of temporal domains, achieving at the order of 50-fold speedup. Simulations of biological processes that undergo long-time evolution, such as oxygen/nutrients transports, thrombus formation/remodeling, or vascular angiogenesis, can benefit most from \AlgAbb{} since the underlying biological processes could take minutes to days.

Lastly, we would like to clarify a misconception about~\AlgAbb{} -- \AlgAbb{} does not predict the path of individual cells. For example, it cannot predict the location of a particular RBC in a future temporal sub-domain given its current location. To expand the statement, \AlgAbb{} cannot predict the distribution of RBCs in the vascular network across time sub-domains. Instead,~\AlgAbb{} is more suitable for simulating well-mixed fluids. In the case of complex fluids such as blood, we are assuming the mixture of plasma and RBCs are homogeneous in the network scope. The inability to track cells across temporal sub-domains is a limitation of \AlgAbb{}. In fact, it is a limitation of parallel-in-time methods for dynamical systems in general. \AlgAbb{} is not suitable to simulate processes that depend on the trajectories of individual particles. 
%These processes usually involve accumulate lasting longer than one temporal sub-domains. In this case, decreasing the number of sub-domains and redirecting the compute resources to SD would be more beneficial.
%\textcolor{blue}{However, the homogeneous assumption is only used when computing flowrate at the beginning and the end of each sub-domain, and it does not impact the integration of the FS inside a sub-domain.}

% \usepackage{appendix}
% \appendix
% \input{Appendix}
\section{Appendix}
\input{Appendix}

\section*{Acknowledgment}
$Tg(kdrl:GFP)$ and $Tg(gata1:DsRed)$ fish lines were provided by D.Y. Stainier at Max Planck Institute for Heart and Lung Research.
The work is partially supported by grant U01 HL142518 from the National Institute of Health.
Y.H. gratefully acknowledges the supports from Fostering Joint International Research (17KK0128) and Grant-in-Aid for Challenging Research (Pioneering: 20K20532) by MEXT (Ministry of Education, Culture, Sports, Science and Technology), Japan.

\bibliographystyle{elsarticle-num}
\bibliography{PaperThree}
\end{document}

%% file: Appendix.tex
\subsection{Dissipative Particle Dynamics}  \label{sec:dpd}
Each DPD particle is represented explicitly by its position and velocity. Its motion is governed by Newton's equations of motion~\cite{Hoogerbrugge1992}:
\begin{align} \label{eqn:Newton}
& \frac{d\mathbf{r}_i}{dr} = v_i ,\\
& \frac{d\mathbf{v}_i}{dt} = \mathbf{F}_i = \sum_{j \neq i} \big(\mathbf{F}^C_{ij} + \mathbf{F}^D_{ij} + \mathbf{F}^R_{ij} \big),
\end{align}
where $t$, $\mathbf{r}_i$, $\mathbf{v}_i$ and $\mathbf{F}_i$ denote time, position, velocity, and force, respectively. The force comprises three components: conservative force $\mathbf{F}^C_{ij}$, dissipative force $\mathbf{F}^D_{ij}$, and corresponding random force $\mathbf{F}^R_{ij}$ from particle $j$ within a radial cutoff $r_c$ of particle $i$. They are expressed as~\cite{Groot1997b}:
\begin{align} 
& \mathbf{F}^C_{ij} = \alpha_{ij} \; \omega_C(r_{ij}) \; \mathbf{e}_{ij}, \label{eqn:FC} \\
& \mathbf{F}^D_{ij} = -\gamma_{ij} \; \omega_D(r_{ij}) \; ( \mathbf{e}_{ij} \cdot \mathbf{v}_{ij} ) \; \mathbf{e}_{ij}, \label{eqn:FD} \\
& \mathbf{F}^R_{ij} = \sigma_{ij} \; \omega_R(r_{ij}) \; \xi_{ij} \; \Delta t_f^{-1/2} \; \mathbf{e}_{ij}, \label{eqn:FR} 
\end{align}
where $\mathbf{e}_{ij}=\mathbf{r}_{ij} / r_{ij}$ is the unit vector between particles $i$ and $j$, and $\mathbf{v}_{ij}$ is the velocity difference. $\Delta t_f$ is the time step of the fine propagator, and $\xi$ is a symmetric Gaussian random variable with zero mean and unit variance~\cite{Groot1997b}. The strengths of the conservative, dissipative, and random force are $\alpha_{ij}$, $\gamma_{ij}$ and $\sigma_{ij}$, respectively. The interaction between pairs of particles are regulated by weighting functions $\omega_C(r_{ij})$, $\omega_D(r_{ij})$ and $\omega_R(r_{ij})$. Most importantly, the balance between dissipation and thermal fluctuation is maintained by the fluctuation-dissipation theorem~\cite{Espanol1995a}:
\begin{equation} \label{FDTforce}
\sigma^2_{ij} = 2 \gamma_{ij} k_B T , \;\;\;\;\;\; \omega_D(r_{ij}) = \omega_R^2(r_{ij}),
\end{equation}
where $k_B T$ is the Boltzmann energy unit. A common choice for the weight functions is $\omega_C(r)=1-r/r_c$ and $\omega_D(r)=\omega^2_R(r)=(1-r/r_c)^2$ for $r\leq r_c$ and zero for $r>r_c$. For simplicity, $k_BT$ and the mass of a particle are taken as the energy unit and mass unit, and their values are set to one.

\subsubsection{RBC and Solvent}
The particle-based blood flow model is based on the DPD approach \cite{Pivkin2008,Fedosov2010a}. In this model, the blood is composed of solvents and red blood cells. The solvent is modeled with single-particles subject to only pairwise forces described in Section \ref{sec:dpd}, and red blood cells are modeled discretely with bonded-particles. Membrane viscosity, elasticity and bending stiffness can be recovered with a spring-network that resembles a triangular mesh on a 2D surface. The shape of the viscoelastic membrane is maintained by a potential derived from a combination of bond, angle and dihedral interactions. The bonds between particles have an attractive and a repulsive component mimicking springs. The attractive potential adopts the form of the wormlike chain potential and is given by \cite{Pivkin2008,Fedosov2010a}
\begin{equation}
V_{WLC} = \frac{k_BTh_m}{4p}\frac{\frac{h}{h_m}^2(3-2\frac{h}{h_m})}{1-\frac{h}{h_m}},
\end{equation}
where $k_BT$ is the energy per unit mass, $h_m$ is the maximum spring extension, and $p$ is the persistence length. The repulsive potential adopts the form of a power function given by
\begin{align}
V_{POW} = \begin{cases} 
	\frac{k_p}{(m-1)h^{m-1}}, & m \neq 1, \\
	-k_p\log(h), & m=1,
	\end{cases}
\end{align}
where $m$ is a positive coefficient, and $k_p$ is force coefficient. Additionally a viscous component damping the springs is realized by a dissipative force, as well as the corresponding random force, given as
\begin{align}
& \mathbf{F}^D_{ij} = -\gamma^T \mathbf{v}_{ij} - \gamma^C (\mathbf{v}_{ij} \cdot \mathbf{e}_{ij}) \mathbf{e}_{ij}, \\
& \mathbf{F}^R_{ij} = \sqrt{2k_BT} \Big(\sqrt{2\gamma^T} (d\mathbf{W}^S_{ij} - tr[d\mathbf{W}^S_{ij}]\mathbf{I}/3) + \frac{\sqrt{3\gamma^C-\gamma^T}}{3} tr[d\mathbf{W}_{ij}] \mathbf{I} \Big),
\end{align}
where $\gamma^T$ and $\gamma^C$ are dissipative coefficients, and $v_{ij}$ is the relative velocity. $d\mathbf{W}^S_{ij}$ is the symmetric component of a random matrix of independent Wiener increments $d\mathbf{W}_{ij}$. 

RBC's structural stability is maintained by the area and volume constraints given by
\begin{align}
& V_{area} = \frac{k_g(A^t-A^t_0)^2}{2A^t_0} + \sum_j \frac{k_l(A_j - A_{0,j})^2}{2A_{0,j}}, \\
& V_{volume} = \frac{k_v(V^t-V^t_0)^2}{2V^t_0},
\end{align}
where $j$ is the triangle index. $k_g$, $k_l$, and $k_v$ are global area, local area, and global volume constraints coefficients, respectively. 
The instantaneous area and volume of a RBC are denoted by $A^t$ and $V^t$, whereas $A^t_0$ and $V^t_0$ represent the equilibrium area and volume. $A^t_0$ is calculated by summing the area of each triangle. $V^t_0$ is found according to scaling relationship $V^t_0/(A^t_0)^{3/2} = V^R / (A^R)^{3/2}$, where $V^R$ and $A^R$ are the experimental measurements. Lastly, the bending resistance of the membrane is modeled by
\begin{equation}
V_{bending} = \sum_j k_b \Big( 1-\cos(\theta_j-\theta_0) \Big),
\end{equation}
where $k_b$ is the bending constant, and $\theta_j$ is the angle between two neighboring triangles with the common edge $j$. We refer interested readers to references \cite{Pivkin2008,Fedosov2010a} for more details. 

%%%%%%%%%%%%%%%%%%%%%%%%%%%%%%%%%%%%%%%%%%%%%%%%
\subsection{1D Blood Flow Model} \label{sec:1Dmodel}

\begin{figure}
    \centering
    \begin{tabular}{c}    
    \subfloat[]{ \includegraphics[width=0.5\textwidth]{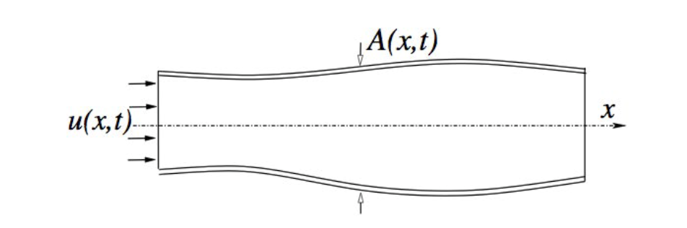} \label{fig:1Dtube} }
    \subfloat[]{ \includegraphics[width=0.5\textwidth]{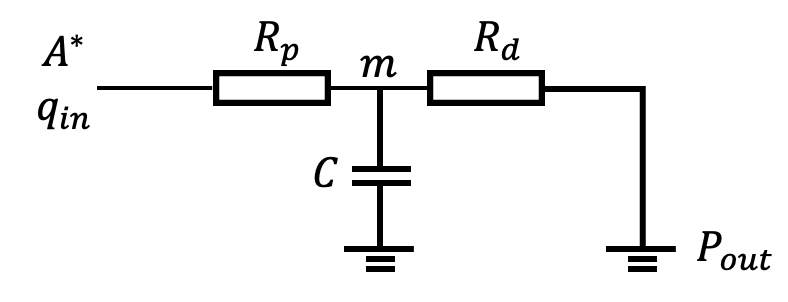} \label{fig:RCR} } 
	\end{tabular}
    \caption{\textbf{(a) Blood vessel as an symmetric tube and (b) a schematic of the RCR boundary condition at outlets for the 1D model.} The assumptions of axial symmetry and radial displacements of the vessel wall reduce the full-order Navier-Stokes equations to an one-dimensional model. This figure is originally illustrated in \cite{Sherwin2003}. }
\end{figure}

Under the assumptions of axial symmetry and radial displacement of the vessel wall (Figure \ref{fig:1Dtube}) \cite{sherwin2003a}\cite{Alastruey2008}, the modeling of blood flow in a compliant vessel can be reduced from full-order Navier-Stokes equations to an 1D model. The 1D model has been widely used for blood flow simulations at a moderate \cite{Sherwin2003} and low \cite{Pindera2009} Reynolds number. In our implementation, it serves as the low-fidelity model supervising the high-fidelity particle model because of its balance between its speed and accuracy.

The 1D model builds on mass conservation and momentum equations \cite{Sherwin2003}:
\begin{equation}
\frac{\partial A}{\partial t} + \frac{\partial AU}{\partial x} = 0
\end{equation}
\begin{equation} \label{eq:1d-mom}
\frac{\partial U}{\partial t} + \frac{1}{2}\frac{\partial U^2}{\partial x} + \frac{1}{\rho}\frac{\partial p}{\partial x} = -\frac{K_rU}{\rho A},
\end{equation}
where $x$ is the axial coordinate along the vessel, and $A(x,t)$ and $p(x,t)$ are the cross-sectional area and intra-luminal pressure, respectively. $K_r$ is the friction loss coefficient representing the viscous resistance of flow per unit length along the vessel. It can be expressed as $K_r = \frac{2\alpha\mu\pi}{\alpha-1}$, where $\alpha = 4/3$ assumes a laminar blood flow with parabolic velocity profile. We assume no tapering effect such as stenosis or atherosclerosis by setting $S = 0$.

The system is closed with Laplace's law on pressure-area:
\begin{equation}
p = p_{ext} + \beta \big( \sqrt[]{A}-\sqrt[]{A_{o}} \big),
\end{equation}
\begin{equation}
\beta = \frac{\sqrt{\pi}h E}{(1-\nu^2)A_{o}},
\end{equation}
The pressure $p$ is the summation of the external pressure $p_{ext}$ and pressure variation, which is characterized by arterial wall compliance $\beta$ and square root of area difference. The Poisson ratio $\nu$ is taken to be 0.5, and $h$ and $A_{o}$ are the vessel wall thickness and reference cross-sectional area. $E$ is the Young's modulus of the vessel wall. 

We use the second-order Adams-Bashforth scheme to discretize and compute the time integration. For spatial discretization, the whole arterial network is decomposed into $\textit{N}$ non-overlapping domains. Each domain can be further subdivided into elements. Within each element, the solution is formulated as a linear combination of orthogonal Legendre polynomials. At element interfaces, the solver uses the discontinuous Galerkin method to render the discontinuous numerical solutions.  We refer interested readers to \cite{sherwin2003a} for details.

The imposed boundary condition must satisfy conservation of mass, which constrains the flow rate at outlets. For the DPD model, we can impose Dirichlet type boundary conditions at inlets and outlets. However, for the 1D system, this would lead to a defective boundary condition~\cite{Formaggia2002} and a problem with well-posedness~\cite{Quarteroni2002}. An alternative to the defective boundary condition is the Windkessel model~\cite{westerhof2009arterial}, which is analogous to open-loop circuits. This approach captures the effective resistance (R) and compliance (C) of the neglected downstream vessels. In particular, we apply a three element "Windkessel" RCR boundary conditions to each of the outlets, as shown in Fig.~\ref{fig:RCR}. The RCR boundary conditions are given as:
\begin{equation}
q_{in} = \frac{P(A^{*}) - P_{m}}{R_{p}} = C\frac{dP_{c}}{dt} + \frac{P_{m} - P_{out}}{R_{d}}
\end{equation}
where $R_{p}$, $R_{d}$, $C$, $A^{*}$ and $q_{in}$ represent proximal and distal resistance, capacitance, area and outflow at 1D outlets. We tune the values of each resistor and capacitor so that the outflow rate and pressure match those of experimental measurements. For more detail, we refer interested readers to \cite{Alastruey2008}.

% \subsection{DPD Parameters} \label{sec:AppendixA}

% For the simple fluid example in Section \ref{sec:SimpleFluid}, the DPD pair interaction parameters in Equation (\ref{eqn:FC},\ref{eqn:FD},\ref{eqn:FR}) are $\alpha=30$, $\gamma=4.5$, $\sigma=3$, $s=2$, $r_\text{c}=1$.

% For the complex fluid example in Section \ref{sec:ComplexFluid}, the DPD pair interaction parameters in Equation (\ref{eqn:FC},\ref{eqn:FD},\ref{eqn:FR}) are tabulated in Table \ref{tab:ComplexPairCoeff}. For the simulation of flow in zebrafish hindbrain (Section \ref{sec:zebrafishhindbrain}), the parameters are tabulated in Table \ref{tab:zebrafishPairCoeff}.

%%%%%%%%%%%%%%%%%%%%%%%%%%%%%%%%%%%%%%%%%%%%%%%%%%%%%%%%%%%%%%%%%%%%%%%%%%%%%%%%%%%%%%%%%%%%%%%%%

\subsection{Mapping to Physical Units} \label{sec:AppendixB}

For the DPD model, we set the length scale $[L^{\text{DPD}}] = 1 \times 10^{-6}$m. This means that one unit length in the DPD system is equivalent to $10^{-6}$ meters of physical length. By matching the viscosity of real blood to that of the DPD fluid, the time scale $[T^{\text{DPD}}]$ can be evaluated with relation:
\begin{equation}
[T^{\text{DPD}}] = [L^{\text{DPD}}] \frac{\nu^\text{P}}{\nu^\text{M}} \frac{\mu^\text{M}}{\mu^\text{P}} (\text{s}),
\end{equation}
where $\mu$ is the RBC membrane shear modulus, and $\nu$ is the viscosity. The superscripts $M$ and $P$ denote the model and physical units, respectively. For the blood simulation, we use the viscosity of plasma $\nu^\text{P} = 1.2 \times 10^{-3}$ Pa s, and the RBC membrane shear modulus $\mu^\text{P}_\text{s} = 4.5 \times 10^{-6}$ Nm$^{-1}$:
\begin{equation}
[T^{\text{DPD}}] = 10^{-6} \frac{1.2 \times 10^{-3}}{21.8132} \frac{100}{4.5 \times 10^{-6}} = 1.2225 \times 10^{-3} \text{s}.
\end{equation}
Here, we set $\mu^\text{M} = 100$ and measured $\nu^\text{M}=21.8132$ from running benchmark simulations.

\begin{figure} \label{fig:para_imple}
\centering
\subfloat[]{\includegraphics[width=0.45\textwidth]{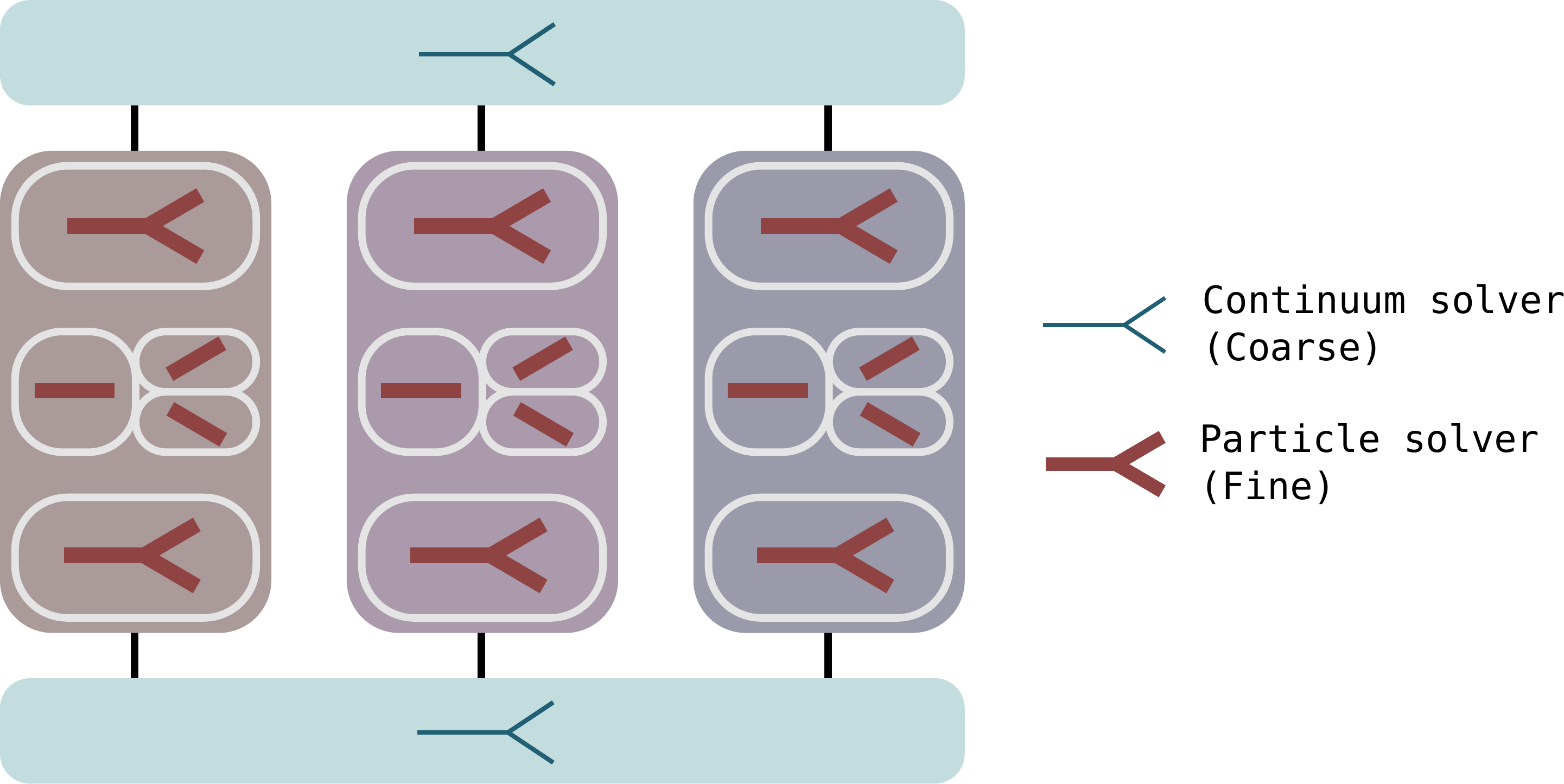}\label{fig:para_imple_1}}
\qquad   % add space in between
\subfloat[]{\includegraphics[width=0.45\textwidth]{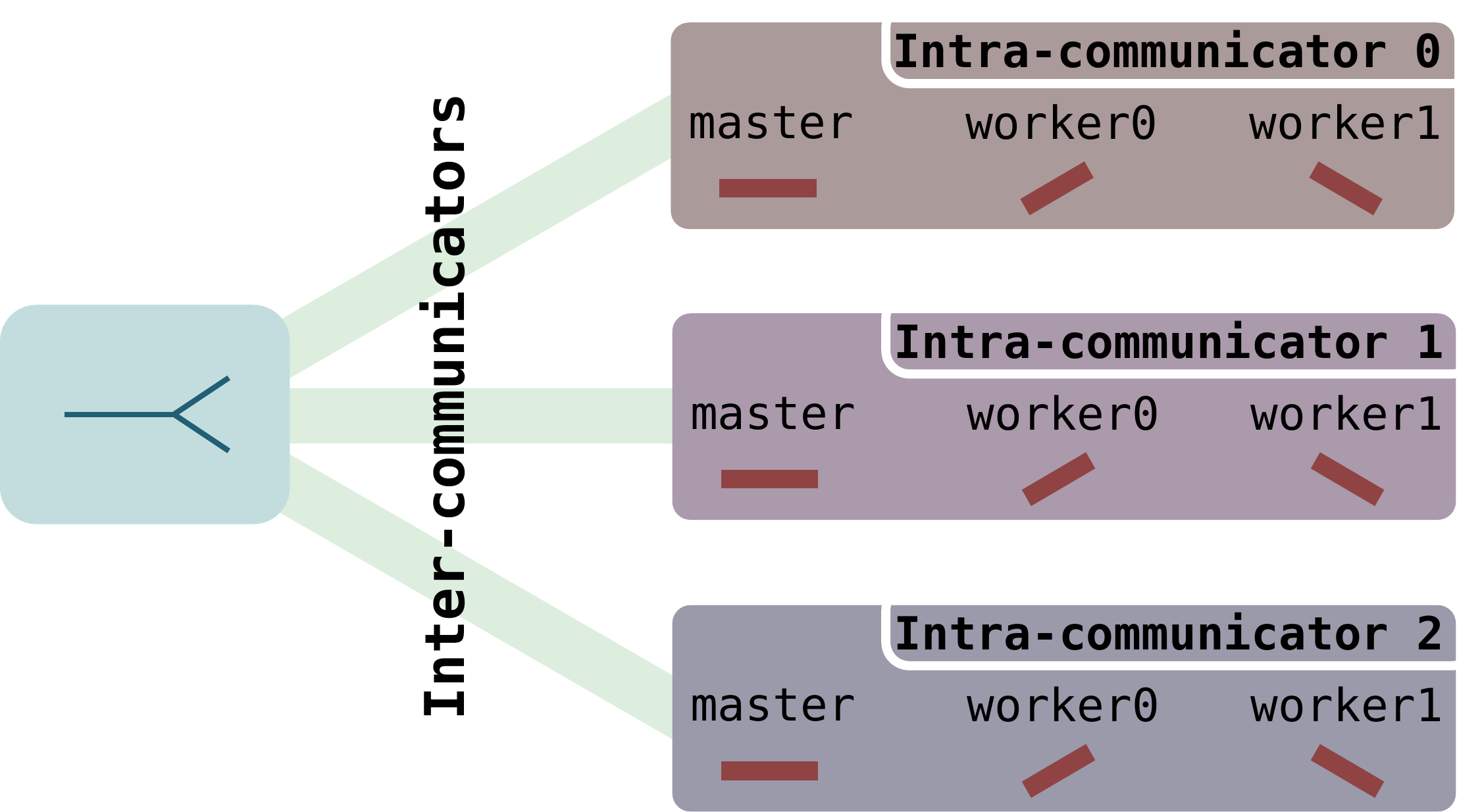}\label{fig:para_imple_2}}	
\caption{\textbf{Pipeline of the parallel implementation.} (a) CS sends the volume-flux to the FSs as initial conditions. The fine-solvers then propagate the systems concurrently. For each FS, its simulation domain is decomposed into multiple sub-domains for additional layer of parallelism. (b) The dual level of parallelism (i.e. spatial and temporal) is achieved owing to the clever arrangement of intra-solver and inter-solver MPI communicators. }
\end{figure}

\subsubsection{Parallel implementation}
The continuum and particle solvers are, by themselves, stand-alone solvers with complex code structures. Implementing codes into the solvers for coupling, while complying with the structure of the software, is a challenging task. To minimize code refactoring, we use Multiscale Universal Interface (MUI). MUI is a library that facilitates the concurrent coupling of heterogeneous solvers \cite{Tang2015}. It integrates MIMD (Multiple instruction streams, multiple data streams) and an asynchronous communication protocol to handle inter-solver information exchange, irrespective of intra-solver MPI implementation. Coupling via MUI does not pollute the MPI communicator space, and therefore requires less code refactoring. 

We illustrate the information-exchange pipeline in Figure \ref{fig:para_imple_1}. In this step, CS sends the volume-flux at various times to the FSs as initial conditions. Each FS receives its initial condition and resets its state to match the condition. Because there is no temporal overlap among the fine-propagations, they can be carried out concurrently. For each FS, its simulation domain is decomposed into multiple sub-domains for spatial acceleration, which creates an additional layer of parallelism. The results are then sent back to the coarse-solver.

The dual level of parallelism (i.e. spatial and temporal) is achieved owing to the clever arrangement of intra-solver and inter-solver communicators. The inter-solver communication is managed by MUI through MPI inter-communicators, which establish links between the CS and each of the FSs, as illustrated in Figure \ref{fig:para_imple_2}. In contrast, inter-solver communication is enabled by MPI intra-communicators. The master of each propagator shares the incoming information among the associated worker processes, each of which handles a sub-domain. At the end of a fine-propagation, the master pools information from the worker processes. The gathered data are then sent back through the inter-solver communicator.